\documentclass[final]{article}

\usepackage{amssymb}
\usepackage{amsfonts}
\usepackage{amsmath}
\usepackage{graphicx}
\usepackage{subfigure}
\usepackage{url}
\usepackage{cite}
\usepackage{hyperref}
\usepackage{showkeys}
\usepackage{color}
\usepackage{algorithm}
\usepackage{algorithmic}
\usepackage{paralist}

\newcommand{\abs}[1]{\lvert#1\rvert}
\newcommand{\norm}[1]{\lVert#1\rVert}
\newcommand{\paren}[1]{\left(#1\right)}
\newcommand{\bracket}[1]{\left[#1\right]}
\newcommand{\set}[1]{\left\{#1\right\}}

\newcommand{\bigo}{\mathrm{O}}

\newcommand{\dx}{\partial_{x} }
\newcommand{\sinc}{\mathrm{sinc}}
\newcommand{\Si}{\mathrm{Si}}

\newcommand{\inner}[2]{\left\langle #1,#2\right\rangle}

\newcommand{\f}{\phi}

\newcommand{\inv}{^{-1}}

\newcommand{\bv}{\bold{v}}

\newcommand{\sech}{\mathrm{sech}}

\newtheorem{thm}{Theorem}[section]

\newtheorem{defn}[thm]{Definition}

\newcommand{\pdiff}[2]{\frac{\partial{#1}}{\partial{#2}}}

\newcommand{\trans}{^{T}}
\newcommand{\Div}{\nabla \cdot}
\newcommand{\Grad}{\nabla}

\newcommand{\Pcmp}{{\cal P}}

\newcommand{\DvD}[3]{\frac{D_{#1}{#2}}{D{#3}}}

\newcommand{\DvDl}[3]{{D_{#1}{#2}/D{#3}}}

\newcommand{\ttarray}[4]{{\setlength{\columnsep}{1.1\columnsep}\left[\begin{array}{cc}{#1} & {#2} \\{#3} & {#4} \\\end{array}\right]}} 
\newcommand{\varray}[2]{{\setlength{\columnsep}{1.1\columnsep}\left[\begin{array}{c}{#1}  \\{#2}  \\\end{array}\right]}}

\begin{document}

\author{G. Simpson\footnote{\href{mailto:simpson@math.toronto.edu}{simpson@math.toronto.edu}}\\
  Department of Mathematics,\\ University of Toronto \and
  M. Spiegelman\footnote{\href{mailto:mspieg@ldeo.columbia.edu}{mspieg@ldeo.columbia.edu}}\\
  Department of Applied Physics \& Applied
  Mathematics,\\
  Department of Earth \& Environmental Sciences, \\ Columbia
  University} 

\title{Solitary Wave Benchmarks in Magma Dynamics}

\maketitle

\abstract{We present a model problem for benchmarking codes that
  investigate magma migration in the Earth's interior.  This system
  retains the essential features of more sophisticated models, yet has
  the advantage of possessing solitary wave solutions.  The existence
  of such exact solutions to the nonlinear problem make it an
  excellent benchmark problem for combinations of solver algorithms.
  In this work, we explore a novel algorithm for computing high
  quality approximations of the solitary waves and use them to
  benchmark a semi-Lagrangian Crank-Nicholson scheme for a finite
  element discretization of the time dependent problem.}

\section{Introduction}
\label{s:overview}

Benchmark problems are of great utility for verifying and comparing
numerical algorithms, and exact solutions play an important role in
constructing such benchmarks.  Unfortunately, exact solutions may be
difficult, or impossible, to construct for nonlinear problems.  In
this work, we formulate a benchmark problem for the simplest
non-linear model of magma migration, and explore algorithms for
constructing the exact solutions and simulating the system.

On the viscous time scale, many processes in the solid Earth,
including mantle convection, magma migration, and crustal deformation,
occur at such low Reynolds numbers that inertial terms can be
neglected.  These {\it quasi-static} systems generically take the form
\begin{subequations}\label{e:quasi_static}
  \begin{align}
    \label{e:hyperbolic}
    \partial_t \psi + \nabla \cdot \paren{\psi {\bf v}} &=
    \text{sources
      and sinks of $\psi$}\\
    \label{e:elliptic}
    \nabla \cdot \sigma({\bf v}; \psi) &= \text{body forces on the
      medium}
  \end{align}
\end{subequations}
where $\psi$ is some material parameter, such as temperature, fluid
fraction or chemical concentration, which influence constitutive
relations such as permeability or rheology of the medium.  Though
\eqref{e:hyperbolic} is written to suggest a hyperbolic nature, it may
of course be parabolic, with diffusive terms appearing as sinks. All
time dependence comes from the first equation, \eqref{e:hyperbolic},
which affects the velocity field through the elliptic problem
\eqref{e:elliptic}, which then advects the scalar fields.  This
coupling introduces its own set of scientific computing challenges;
there is a simultaneous need for a robust, fast, elliptic solver and
an efficient time stepping algorithm.


\subsection{A Reduced Model for Magma Migration}
An important example of a coupled hyperbolic-elliptic systems in Earth
science is the PDE's governing the flow of a low viscosity fluid in a
viscously deformable porous matrix which has been used to model the
flow of partially molten rock (magma) in the Earth's interior.
Beginning with the primitive equations first formulated in
\cite{mckenzie1984gac}, one can, after many simplifications, arrive at
the system:
\begin{subequations}\label{e:magma_system}
  \begin{align}
    \phi_t &= \phi^m \mathcal{P},\\
    \bracket{\phi^m - \nabla\paren{\phi^n\nabla \cdot}} \mathcal{P} &
    = -\nabla\cdot {\phi^n{\bf e}_d}.
  \end{align}
\end{subequations}
In the above equations: \begin{inparaenum}[\itshape (i)] \item $\phi$
  is porosity, or volume fraction of melt; \item $\Pcmp$ is the
  ``compaction pressure'', measuring viscous deformation associated
  with a compaction of the rock; \item ${\bf e}_d$ is the the unit
  vector in coordinate $d$ associated with the direction of
  gravity\end{inparaenum}.  The above, dimensionless, Eqs.\
\eqref{e:magma_system} have been scaled as follows.  Porosity is
scaled to a reference value $\phi_0\sim0.1-1$\% and distance to the
\emph{compaction length}, an intrinsic length scale of the primitive
system \cite{mckenzie1984gac}.  The compaction length depends on
porosity, but we scale to $\delta_0$, the compaction length at the reference
porosity. All computations here are performed with
respect to the dimensionless equations, however, we will often refer to lengths
 in terms of the number of compaction lengths, multiples of $\delta_0$.


Eq.\ \eqref{e:magma_system}, as written, is more amenable to adding
additional physics such as solid advection or melting. However, it can
also be formulated as a single equation, together with a far field
boundary condition as:
\begin{equation}
  \label{eq:magma}
  \f_t + \partial_{x_d}\paren{\f^n} - \nabla\cdot \bracket{\f^n
    \nabla\paren{\f^{-m}\f_t}}=0, \quad \lim_{\abs{\mathbf x}\to
    \infty} \f(\mathbf x, t) = 1.
\end{equation}
Derivations from primitive equations for conservation of mass,
momentum, and energy are given in \cite{scott1984ms, barcilon1986nwc,
  Spiegelman1993a}.

The important feature of \eqref{eq:magma} is that it possesses {\it
  solitary wave} solutions, localized states which propagate in space
at a fixed speed without changing shape.  Such solutions are ideal for
benchmarking as one need only check the distortion of the transported
waveform.  Indeed, in a frame moving with the solitary wave, the
solution will appear constant.  Numerical studies of the equation and
its solitary waves have been performed in one \cite{barcilon1986nwc,
  scott1984ms}, two \cite{scott1986map}, and three
\cite{wiggins1995mma} dimensions.  Recent work
\cite{Simpson07,simpson08as,simpson08hpde} has shown the
well-posedness of \eqref{eq:magma} and the stability of its solitary
waves in dimension $d=1$.

We demonstrate:
\begin{enumerate}
\item A novel algorithm for computing the solitary waves, which solve
  a time independent equation, based on the Cardinal Whittaker sinc
  function which provide better than polynomial accuracy,
\item A semi-Lagrangian Crank-Nicolson algorithm for a finite element
  discretization of \eqref{e:magma_system}, as one algorithm for the
  benchmarking.
\end{enumerate}

An outline of this paper is as follows.  In section
\ref{s:solwave_properties} we review some properties of the solitary
waves.  We then discuss the sinc collocation method in section
\ref{sec:sinc_disc}, and how it can be implemented for these
equations.  Then, in section \ref{sec:computations}, we demonstrate
the the algorithm with convergence results.  Section \ref{s:time}
explains the time stepping algorithm and its performance.  We offer
some remarks and comments in section \ref{s:discussion}.

\section{Solitary Wave Solutions}
\label{s:solwave_properties}
The solitary wave solutions of \eqref{eq:magma} are exponentially
decaying, radially symmetric humps in excess of $\phi=1$, traveling in
the $x_d$ direction at a fixed speed.  They are akin to the soliton
solutions of the Zakharov-Kuznetsov equation,
\cite{Zakharov:1974p159}, found in plasma physics.

Making the {ansatz}
\begin{equation}
  \label{eq:solwave_ansatz}
  \f(\bold{x},t)= \f_c\paren{\sqrt{\sum_{j=1}^{d-1} x_j^2+(x_d-ct)^2}}= \f_c(r),
\end{equation}
the solitary waves solve the third order equation
\begin{equation}
  \label{eq:solwave_ode}
  0=\begin{cases}
    \begin{split}
      -c \f_c ' &+ \paren{\f_c^n}' + \frac{c}{1-m}\paren{\f_c^n
        (\f_c^{1-m})''}' \\
      &+ \frac{c (d-1)}{1-m}\f_c^n \paren{\frac{1}{r}
        (\f_c^{1-m})'}'\end{split}& m \neq 1\\
    \begin{split}-c \f_c ' &+ \paren{\f_c^n}' + c\paren{\f_c^n
        (\log\f_c)''}' \\
      &+ c (d-1)\f_c^n \paren{\frac{1}{r} (\log\f_c)'}'\end{split}& m
    = 1
  \end{cases}
\end{equation}
with the boundary conditions
\begin{subequations}
  \label{eq:solwave_bc}
  \begin{align}
    \f_c'(0)&=0\\
    \lim_{r\to\infty} \f_c(r) &= 1
  \end{align}
\end{subequations}
Derivatives are taken with respect to $r$.

The existence of solitary wave solutions was proven by a phase plane
argument in dimension one.  Further smoothness properties were stated
in \cite{simpson08as} as part of the stability analysis .  The formal
existence of solitary wave solutions in higher dimensions is an open
problem, though it is expected.

In one dimension, \eqref{eq:solwave_ode} can always be integrated up
twice, reducing the problem to quadrature.  In the case case $n=3$ and
$m=0$, after one integration we have
\begin{equation}
  \label{eq:solwave_1dim}
  -c(\f_c-1) + \f_c^3-1 + c\f_c^3 \f_c'' =0
\end{equation}
Two more integrations give the implicit expression
\begin{equation}
  \label{eq:solwave_implicit}
  r^2
  = \paren{A+\frac{1}{2}}\bracket{-2\sqrt{A-\f_c}+\frac{1}{\sqrt{A-1}}\log\paren{\frac{\sqrt{A-1}-\sqrt{A-\f_c}}{\sqrt{A-1}+\sqrt{A-\f_c}},
    }}^2
\end{equation}
where $A = (c-1)/2$ is the amplitude.

\subsection{Reformulation by Even Extension}
For higher dimensions, such integrations of \eqref{eq:solwave_ode} are
not possible and we resort to numerical approximation.  First,
\eqref{eq:solwave_ode} is rewritten to make it more amenable to
computation.  Integrating from $r$ to $\infty$ yields the
integro-differential equation
\begin{equation}
  \label{eq:solwave_integro}
  0=\begin{cases}
    \begin{split}-c (\phi_c -1) &+ \paren{\f_c^n-1} +
      \frac{c}{1-m}\paren{\f_c^n
        (\f_c^{1-m})''} \\
      &-\frac{c (d-1)}{1-m}\int_r^\infty\f_c^n \paren{\frac{1}{r}
        (\f_c^{1-m})'}'dr \end{split}& m \neq 1\\
    \begin{split}-c (\phi_c -1) &+ \paren{\f_c^n-1} + c\paren{\f_c^n
        (\log\f_c)''} \\
      &-c (d-1)\int_r^\infty \f_c^n \paren{\frac{1}{r} (\log\f_c)'}'
      dr \end{split}& m = 1
  \end{cases}
\end{equation}
The sinc-collocation method we wish to apply does not apply directly
to problems posed on the semi-axis, $(0,\infty)$.  The problem must be
altered to live on all of $(-\infty, \infty)$. Let
\begin{equation}
  \label{eq:solwave_even}
  \tilde\phi_c(x) = \f_c(\abs{x})
\end{equation}
be the even extension of $\f_c$ to the whole real line. Dropping the
$\tilde{\;}$'s, $\phi_c$ solves
\begin{equation}
  \label{eq:solwave_integro_e}
  0=\begin{cases}
    \begin{split}-c(\phi_c-1) &+{\phi_c^n-1} +
      \frac{c}{1-m}\paren{\phi^n (\phi_c^{1-m})''} \\
      &+\frac{c (d-1)}{1-m} \int_{-\infty}^x\phi_c^n \paren{\frac{1}{x} (\phi_c^{1-m})'}'dx  \end{split}& m \neq 1\\
    \begin{split}-c(\phi_c-1) &+ \phi_c^n-1 + c\paren{\phi_c^n
        (\log\phi_c)''} \\
      &+ c (d-1) \int_{-\infty}^x\phi_c^n \paren{\frac{1}{x} (\log
        \phi_c)'}' dx \end{split}& m = 1
  \end{cases}
\end{equation}
We discretize and solve \eqref{eq:solwave_integro_e} using the
sinc-collocation method, described in Section \ref{sec:sinc_disc}.





\section{Collocation \& Continuation}
\label{s:colloc_cont}
Equation \eqref{eq:solwave_integro_e} is a nonlinear
integro-differential equation posed on an unbounded domain.  We employ
a method, sinc-collocation, that respects these features. The sinc
spectral method is thoroughly formulated and explained in
\cite{lund1992smq, stenger1993nmb, stenger1981nmb, stenger2000ssn} and
briefly in Appendix \ref{app:sinc_details}.  In the sinc
discretization, the problem remains posed on $\mathbb{R}$ and the
boundary conditions, that the solution vanish at $\pm \infty$, are
naturally incorporated.  This method has been applied to a variety of
differential equations; see the above references.  It has been used to
compute solitary waves in the early work \cite{Lundin:1980p356} and
more recently in \cite{Marzuola:2010p5455}.  It was also used to study
a dependent problem, KdV, in \cite{alkhaled2001sns}.

Collocation insists the equation be satisfied at the nodes of the
mesh.  This is in contrast to a Galerkin formulation, which would have
us discretely orthogonalize the residual against some family of
functions.  Collocation can be interpreted as discretely solving the
classical form of the equation, while Galerkin discretely solves the
weak form. Moshen \& El-Gamel found collocation to be superior for
certain problems, \cite{mohsen2008gac}.

This discretization will lead to a nonlinear system of algebraic
equations, requiring a good initial guess for the solver.  Our strategy
for constructing such a solution is to take the $d=1$ solution and
then perform numerical continuation in dimension, up to the desired
value.  Constructing the $d=1$ solution is also done by continuation,
using an asymptotic approximation in the small amplitude, $c \sim n$,
state; continuation is performed in $c$ to its desired value.


\subsection{Sinc Discretization}
\label{sec:sinc_disc}

Given a function $u:\mathbb{R} \to \mathbb{R}$, $u$ is approximated
using a superposition of shifted and scaled $\sinc$ functions:
\begin{equation}
  \label{eq:sinc_approx}
  C_{M, N}(u,h)(x) \equiv \sum_{k = -M}^N u_k
  \sinc\paren{\frac{x-x_k}{h}} = \sum_{k = -M}^N u_k S(k,h) (x), 
\end{equation}
where $x_k = k h$ for $k = -M, \ldots, N$ are the \emph{nodes} and
$h>0$.  There are \emph{three} parameters in this discretization, $h$,
$M$, and $N$, determining the number and spacing of the lattice
points.  This is common to numerical methods posed on unbounded
domains, \cite{boyd2001caf}.

A useful and important feature of this spectral method is that the
$S(k,h)$ functions act like discrete delta functions,
\begin{equation}
  \label{eq:disc_delta}
  C_{M,N}(u,h)(x_k) = u_k.
\end{equation}
For sufficiently smooth functions, the convergence of this
approximation is rapid both in practice and theoretically.  See
Theorem \ref{thm:sinc_convergence} in Appendix \ref{app:sinc_details}
for a statement on optimal convergence.

Since the solution is even, we may take $N = M$; we write
\begin{equation}
  C_{M}(u,h)(x) \equiv C_{M,M}(u,h)(x) .
\end{equation}

We further reduce the number of free parameters down to just $M$, by
slaving $h$ to $M$ as
\begin{equation}
  \label{e:h_relation}
  h = \pi\sqrt{\frac{1}{2 \gamma M}}, \quad \gamma = \sqrt{1 - \frac{n}{c}}
\end{equation}
The motivation for this choice is discussed in Appendix
\ref{app:sinc_details}.  It is closely connected to the theory of the
sinc method, and the asymptotic decay properties of both $\phi_c-1$
and its Fourier transform.

Letting $u_c = \phi_c-1$, the solitary wave with the asymptotic state
subtracted off, we formulate \eqref{eq:solwave_integro_e} as a
nonlinear collocation problem at the nodes $\set{x_k}$:
\begin{equation}{
  0=\begin{cases}
    \begin{split}
      -c C_{M}(u_c,h)(x_k) + \paren{C_{M}(u_c,h)(x_k) +1}^n - 1 \\
      \quad+ \frac{c}{1-m} \paren{C_{M}(u_c,h)(x_k) +1}^n \frac{d^2}{dx^2} \paren{C_{M}(u_c,h)(x_k)}^{1-m}  \\
      + \frac{ c (d-1)}{m-1} \int_{x_k}^\infty
      (C_{M}(u_c,h)(x)+1)^n{x} \frac{d}{dx}\set{\frac{1}{x}
        \frac{d}{dx}\bracket{\paren{C_{M}(u_c,h) (x)}^{1-m}}} dx.
    \end{split} & m \neq 1\\
    \begin{split}
      -c C_{M}(u_c,h)(x_k) + \paren{C_{M}(u_c,h)(x_k) +1}^n - 1 \\
      \quad+ c \paren{C_{M}(u_c,h)(x_k) +1}^n \frac{d^2}{dx^2} \paren{\log C_{M}(u_c,h)(x_k)}  \\
      + c (d-1) \int_{x_k}^\infty (C_{M}(u_c,h)(x)+1)^n{x}
      \frac{d}{dx}\set{\frac{1}{x}
        \frac{d}{dx}\bracket{\log{C_{M}(u_c,h) (x)}}} dx.
    \end{split} & m = 1
  \end{cases}}
\end{equation}
{\bf From here on we suppress the subscripts $c$ in $u_c$ and
  $\phi_c$.} 

Let ${\bf u}$ be the column vector associated with the sinc
discretization of $u$, at the collocation points $\set{x_k}$,
\begin{equation}
  C_M(u,h)(x_k) \mapsto {\bf u} = \begin{pmatrix} u_{-M} &  u_{-M+1} &
    \ldots &  u_M \end{pmatrix}^T .
\end{equation}
Associated with ${\bf u}$ is $\boldsymbol{\phi} = {\bf u + 1}$.

We now define a series of matrices that operate on ${\bf u}$.  The
derivatives of sinc approximated functions are given by:
\begin{equation}
  \label{eq:matrix_dl}
  D^{(l)}_{jk} = \frac{d^l}{dx^l}S(j,h)(x)|_{x = x_k}
\end{equation}
Explicitly,
\begin{subequations}
  \begin{align}
    D^{(1)}_{jk} & = \begin{cases}0 & j =k \\
      \frac{1}{h} \frac{(-1)^{k-j}}{k-j} & j \neq k
    \end{cases}\\
    D^{(2)}_{jk} &= \begin{cases} \frac{1}{h^2} \frac{-\pi^2}{3} & j=k\\
      \frac{1}{h^2} \frac{-2 (-1)^{k-j}}{(k-j)^2} & j \neq k
    \end{cases}
  \end{align}
\end{subequations}
The integration matrix is
\begin{equation}
  D^{(-1)}_{jk} = \frac{h}{2}+ \frac{h}{\pi}\Si(\pi(j-k))
\end{equation}
where $\Si$ is the sine-integral function,
\begin{equation}
  \Si(x) \equiv \int_0^x \frac{\sin(t)}{t}dt.
\end{equation}

Being singular, the $\frac{1}{x} \frac{d}{dx}$ operator must be
treated carefully.  For smooth even functions ($u'(0) = 0$), it is
well defined.  Taking limits of the sinc approximation of an even
function ($u_{-k} = u_k$),
\begin{align}
  \lim_{x\to 0} \frac{1}{x}\dx u_0 S(0,h)(x) &= -\frac{\pi^2}{3 h^2}u_0,\\
  \lim_{x\to 0}\frac{1}{x} \paren{\dx u_kS(k,h)(x) +\dx
    u_{-k}S(-k,h)(x)}&= -4 \frac{(-1)^k}{h^2 k^2}u_k.
\end{align}
The matrix $\tilde{D}^{(1)}$ approximating $\frac{1}{x}\frac{d}{dx}$
is defined as:
\begin{equation}
  \tilde{D}^{(1)}_{jk} = \begin{cases}
    \frac{1}{x_j}D^{(1)}_{jk} & j \neq 0,\\
    -\frac{2(-1)^k}{h^2 k^2} & j = 0, k \neq 0,\\
    \frac{-\pi^2}{3h^2} & j = k = 0.
  \end{cases}
\end{equation}

With these matrices, the discretization of
\eqref{eq:solwave_integro_e} is equivalent to the nonlinear algebraic
system
\begin{equation}
  \label{eq:solwave_colloc}
  {\bf F}({\bf u}) =\begin{cases} 
    \begin{split}-c {\bf u} &+ \boldsymbol{\phi}^n - {\bf 1}+
      \frac{c}{1-m}\boldsymbol{\phi}^n D^{(2)} ({
        \boldsymbol{\phi}}^{1-m}- {\bf 1}) \\
      &+ \frac{c (d-1)}{1-m} D^{(-1)} (\boldsymbol{\phi})^n
      D^{(1)}\tilde{D}^{(1)} ({\boldsymbol{\phi}}^{1-m}- {\bf
        1}) \end{split} & m \neq 1 \\
    \begin{split} - c {\bf u} &+ \boldsymbol{\phi}^n - {\bf 1} + c
      \boldsymbol{\phi}^n D^{(2)} \log \boldsymbol{\phi} \\
      &+ c (d-1) D^{(-1)} \boldsymbol{\phi}^n D^{(1)} \tilde{D}^{(1)}
      \log \boldsymbol{\phi}\end{split}& m = 1
  \end{cases}
\end{equation}
where ${\bf 1}$ is a vector of size $2M+1$ with 1's in all entries.
Nonlinear terms should be interpreted as component-wise operations on
the vectors.

\subsection{Initial Guesses and Numerical Continuation}
To solve \eqref{eq:solwave_colloc}, one needs a good initial guess of
${\bf u}$.  For dimension one, an excellent guess is available.
Integrating \eqref{eq:solwave_ode} reduces the equation to first
order, which can be solved by quadrature and root finding.  Sometimes
it is even possible to obtain implicit solution, as in
\eqref{eq:solwave_implicit}.  This is not possible for $d>1$, nor is
it always desirable to work out the quadrature formulas.  Thus, for a
given $c$ and $d$, we proceed in two steps:
\begin{itemize}
\item For $d=1$, perform numerical continuation in $c$, from a value
  of $c \sim n$, up to the desired value.
\item For $d>1$, apply the continuation in $c$ to construct the
  solution in $d=1$, then perform numerical condition in $d$, up to
  the desired dimension.
\end{itemize}

\subsubsection{Continuation in $c$}
In \cite{Whitehead:1986p71}, the authors observed that in the limit of
small amplitude disturbances of the reference state, \eqref{eq:magma}
was, to leading order, governed by the Korteweg - de Vries equation.
Generalizing this observation in \cite{simpson08as}, let
\begin{subequations}
  \begin{align}
    \gamma &= \sqrt{1 - \frac{n}{c}},\\
    \phi_c(x_1-ct) & = 1 + \frac{\gamma^2}{n-1} U(\gamma (x_1- ct)).
  \end{align}
\end{subequations}
Then $U$ solves the equation
\begin{equation}
  - U + \frac{1}{2}U^2 + \partial_\xi^2 U = \bigo(\gamma^2),
\end{equation}
and small amplitude solitons, where $0 < \gamma \ll 1$, are
approximately
\begin{equation}
  \label{e:small_amp_asympt}
  \phi_c(r)  = 1 + \frac{3\gamma^2}{n-1} \sech^2\left(
    \frac{1}{2}\gamma r\right) + \bigo(\gamma^4).
\end{equation}

Given the desired value of $c$, we partition $(n,c]$ into $P$ points
\begin{equation}
  n < c_1 < c_2 < \ldots < c_P = c
\end{equation}
we iteratively solve
\begin{equation}
  {\bf G}({\bf u};c) =\begin{cases}- c {\bf u} + \boldsymbol{\phi}^n -
    {\bf 1} + \frac{c}{1-m} \boldsymbol{\phi}^n D^{(2)} ({\bf
      \boldsymbol{\phi}}^{1-m}- {\bf 1})& m \neq 1 \\
    - c {\bf u} + \boldsymbol{\phi}^n - {\bf 1} + \frac{c}{1-m} \boldsymbol{\phi}^n D^{(2)} \log \boldsymbol{\phi} & m = 1 
  \end{cases}
\end{equation}
using ${\bf u}^{(j)}$ as the initial guess for
\begin{equation}
  {\bf G}({\bf u}^{(j+1)};c_{j+1}) = {\bf 0}
\end{equation}
The ${\bf u}^{(1)}$ guess is given by \eqref{e:small_amp_asympt}.  We
have successfully solved with $P =\bigo\left( 10\tfrac{c}{n}\right)$,
though this could likely be refined with more sophisticated
continuation algorithms.

\subsubsection{Continuation in $d$}
For $d>1$ we solve by numerical continuation in {dimension} by making
$d$ a parameter:
\begin{equation}
  {\bf H}({\bf u};d) =\begin{cases} \begin{split}-c {\bf u} &+ \boldsymbol{\phi}^n -
      {\bf 1} + \frac{c}{1-m} \boldsymbol{\phi}^n D^{(2)}({\bf
        \boldsymbol{\phi}}^{1-m}- {\bf 1}) \\
      &+ \frac{c (d-1)}{1-m} D^{(-1)} \boldsymbol{\phi}^n
      D^{(1)} \tilde{D}^{(1)} ({\bf
        \boldsymbol{\phi}}^{1-m}- {\bf 1}) \end{split}& m \neq 1 \\
    \begin{split}- c {\bf u} &+ \boldsymbol{\phi}^n - {\bf 1} + c
      \boldsymbol{\phi}^n D^{(2)} \log \boldsymbol{\phi} \\
      &+c(d-1) D^{(-1)} \boldsymbol{\phi}^n D^{(1)} \tilde{D}^{(1)}
      \log \boldsymbol{\phi} \end{split}& m = 1
  \end{cases}
\end{equation}
Given the dimension $d$ for which we desire a solution, we partition
$[1, d]$ into
\begin{equation}
  1 = d_0 < d_1< d_2 < \ldots d_P = d.
\end{equation}
Then, assuming we have solved
\[ {\bf H} ({\bf u}^{(j)}; d_j) = {\bf 0},
\]
${\bf u}^{(j)}$ becomes the initial guess for $d_{j+1}$.  $P$, the
partition size of $[1,d]$, need not be that large.  $P = \bigo(10 d)$
appears sufficient.  As with continuation in $c$, more sophisticated
continuation algorithms might improve this.

\subsection{System Size Reductions}
The even symmetry can be exploited to reduce the size of the algebraic
system.  Since $u_{-k} = u_k$, we need only track $u_k$, $k=0, 1,
\ldots M$.  The symmetry is imposed on \eqref{eq:solwave_colloc} by
the following manipulations on a discretized operator, $A$.  Since
only the last $M+1$ rows are required, we only retain $A_{ij}$ for $i
= M+1,\ldots 2M+1$.  Next, we add or subtract the columns $A_{ij}$, $j
= 1, \ldots M$ onto the columns $j = 2 M+1, \ldots M+2$.  For even/odd
symmetry preserving operations, $\frac{d^2}{dx^2}$ and
$\frac{1}{x}\frac{d}{dx}$, we add.  For even/odd symmetry reversing
operations, $\frac{d}{dx}$ and $\int_{-\infty}^x$, we subtract.  This
reduces the system to $M+1$ points.




\section{Example Solitary Wave Computations}
\label{sec:computations}

We implemented our algorithm using NumPy/SciPy.  The codes for
computing the sinc-collocation matrices were motivated by the {\sc
  Matlab} codes discussed in \cite{weideman2000mdm}.

\subsection{Solitary Wave Forms}
\begin{figure}
  \centering
  \subfigure{\includegraphics[width=2.2in]{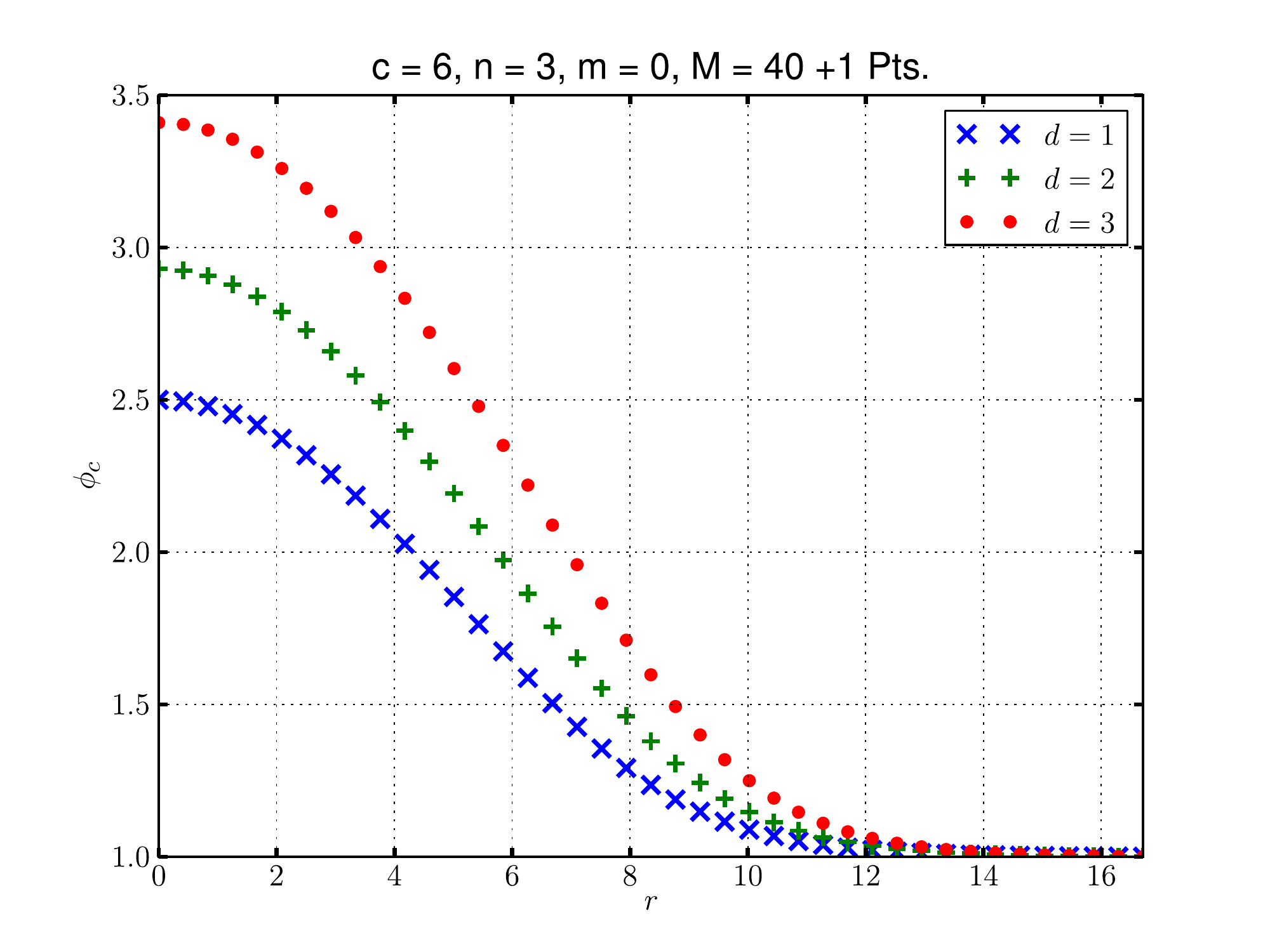}}
  \subfigure{\includegraphics[width=2.2in]{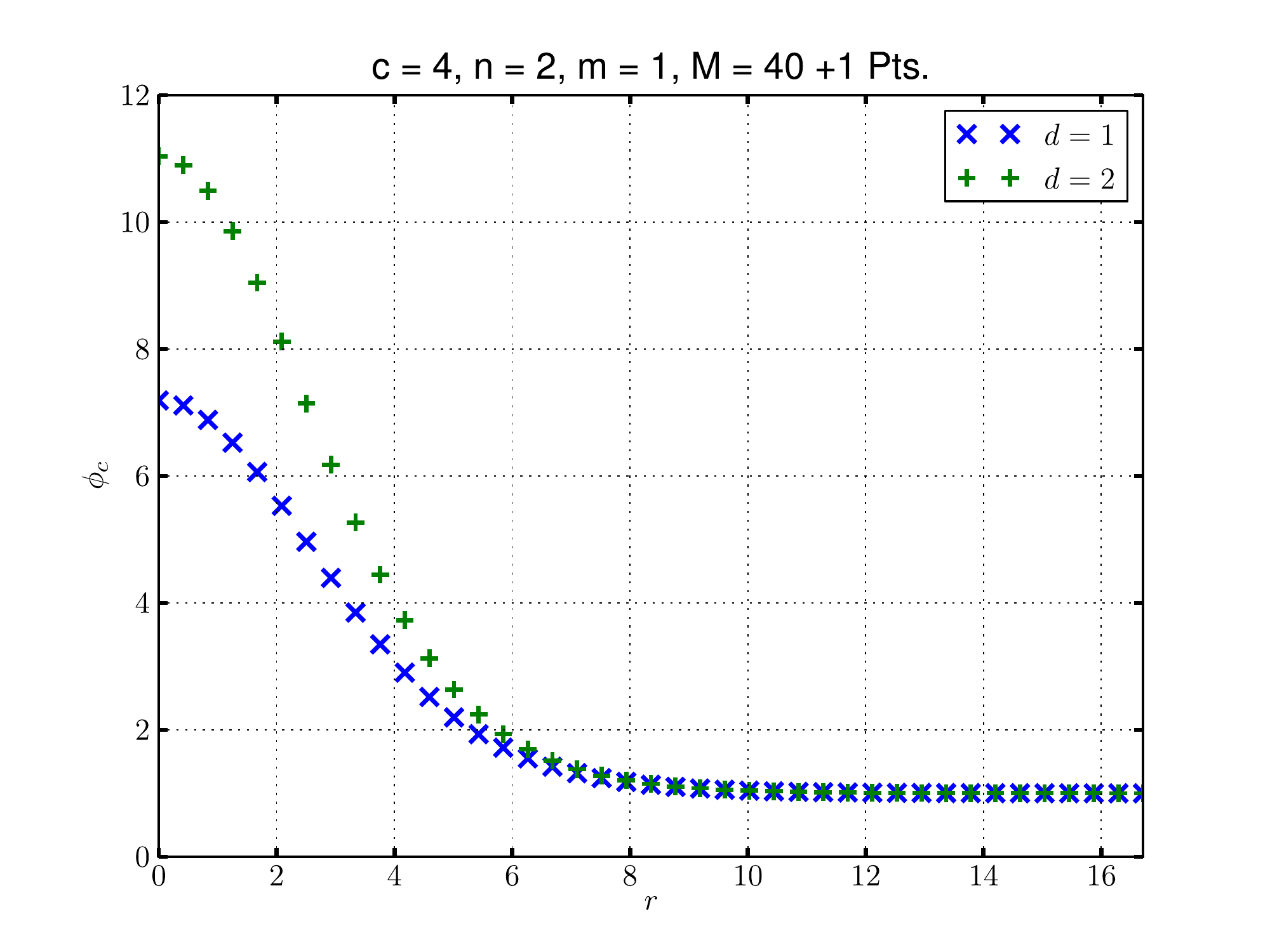}}

  \subfigure{\includegraphics[width=2.2in]{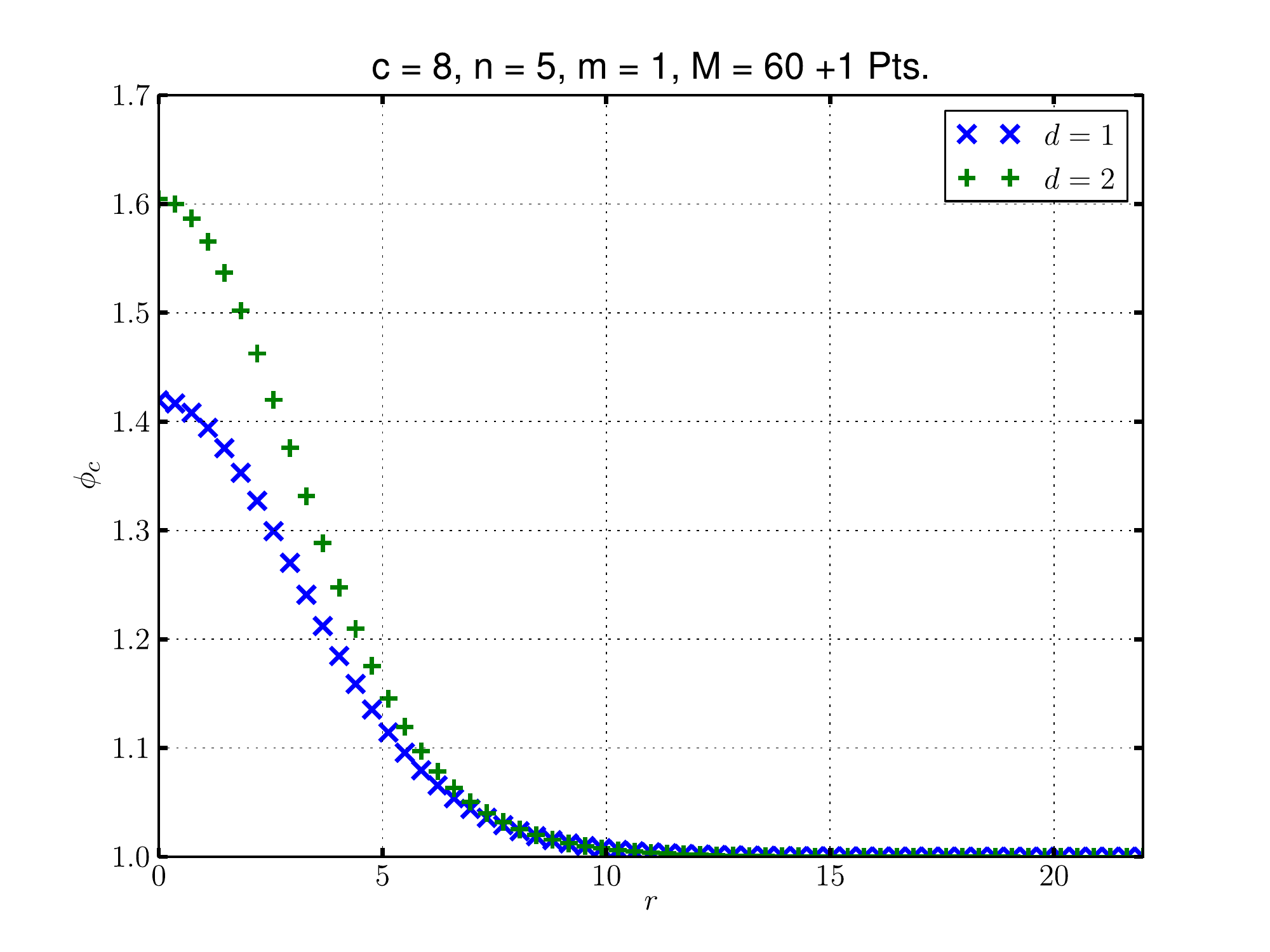}}
  \subfigure{\includegraphics[width=2.2in]{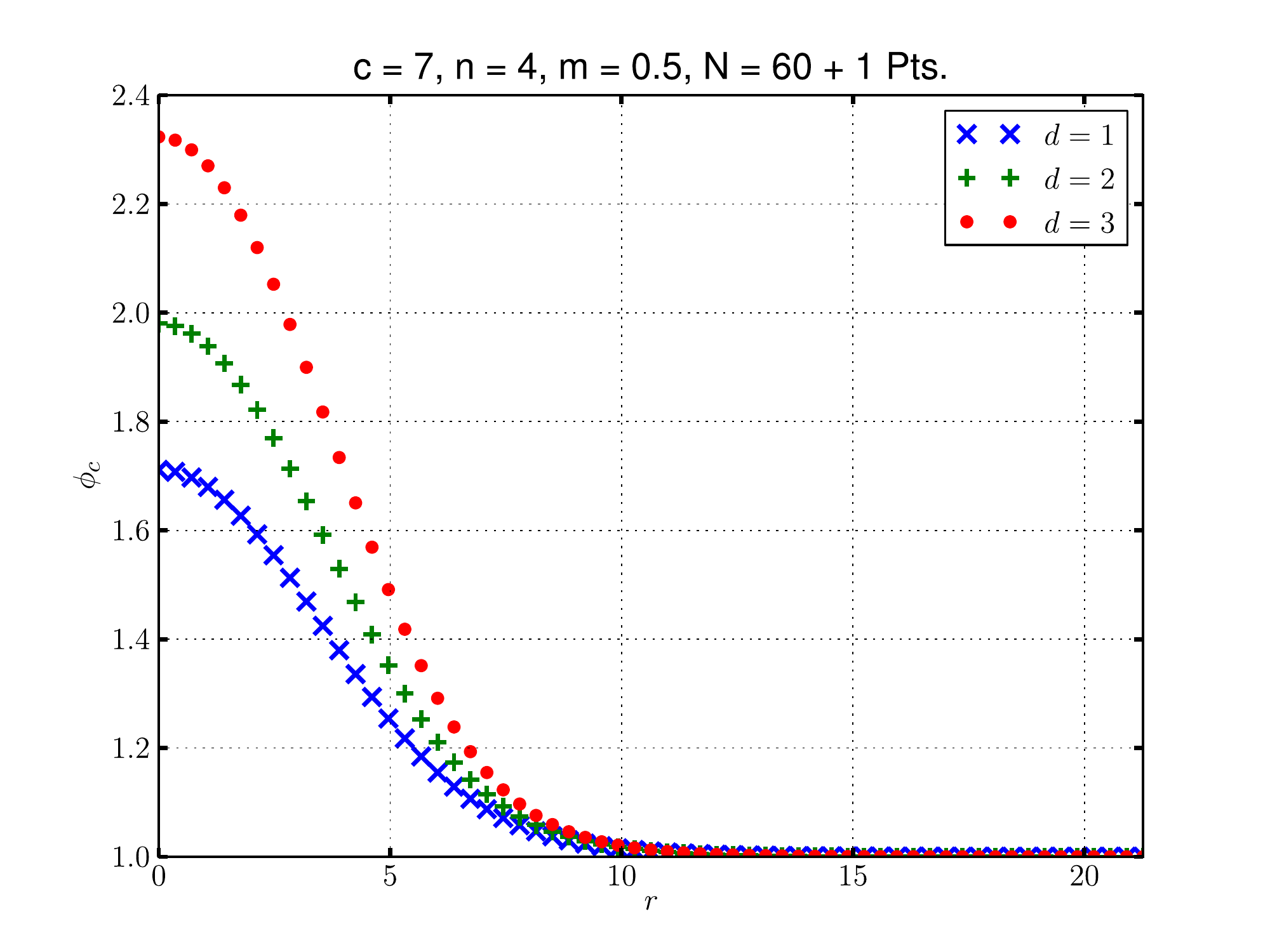}}
  \caption{Examples of computed solitary waves with different values
    of $c$, $n$, $m$, and $M$.}
  \label{f:soliton_examples}
\end{figure}

As a first example of our results, we compute a collection of solitary
waves for different parameter values and dimensions.  These wave forms
are pictured in Figure \ref{f:soliton_examples}.  The amplitude of the
wave tends to increase with dimension.  We have not observed a choice
of $(c,n,m)$ for which this does not happen.  This observation was
previously noted in \cite{wiggins1995mma} for the $n=3$, $m=0$ case.

\begin{figure}
  \centering
  \subfigure{\includegraphics[width=2.2in]{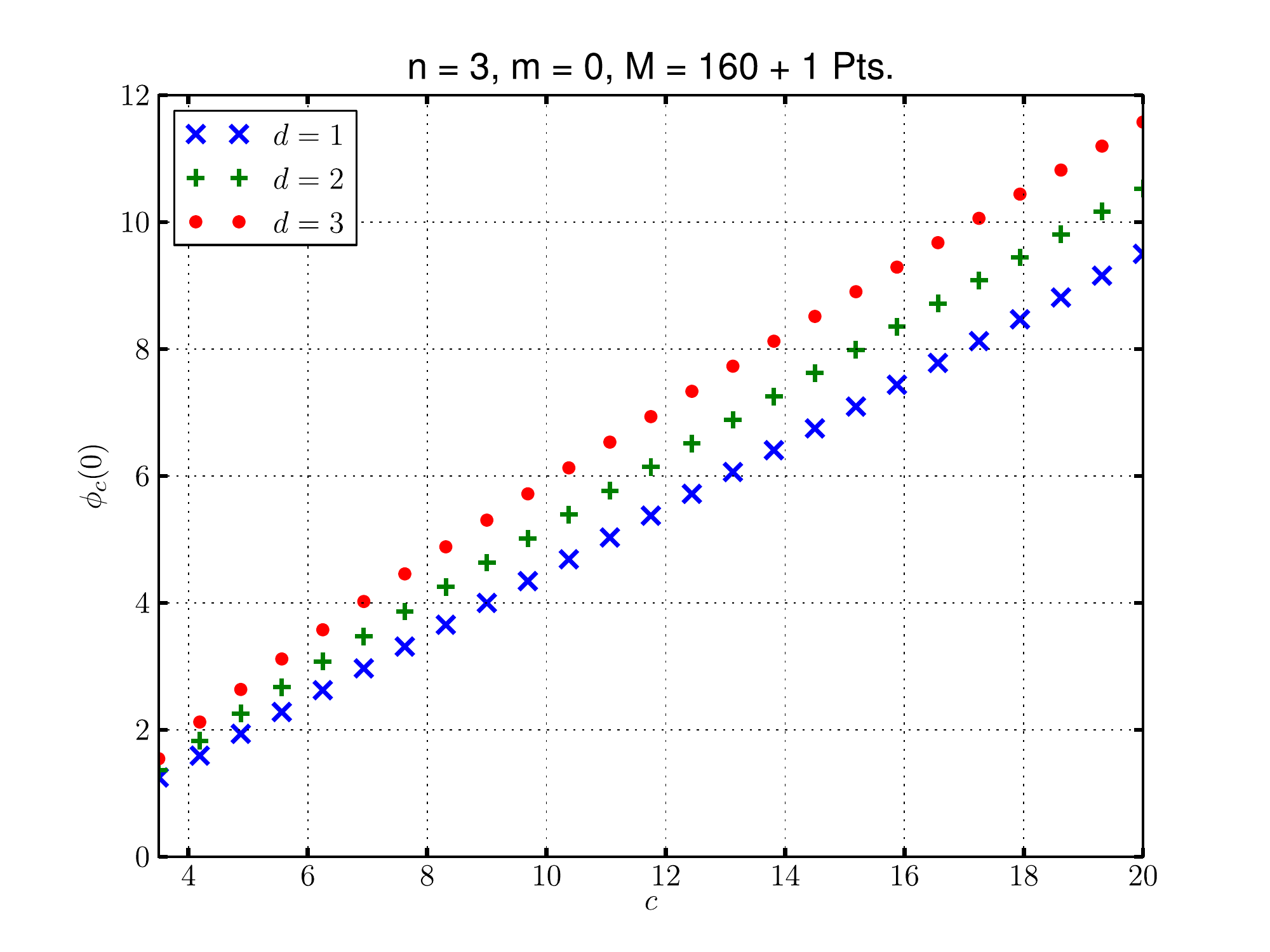}}
  \subfigure{\includegraphics[width=2.2in]{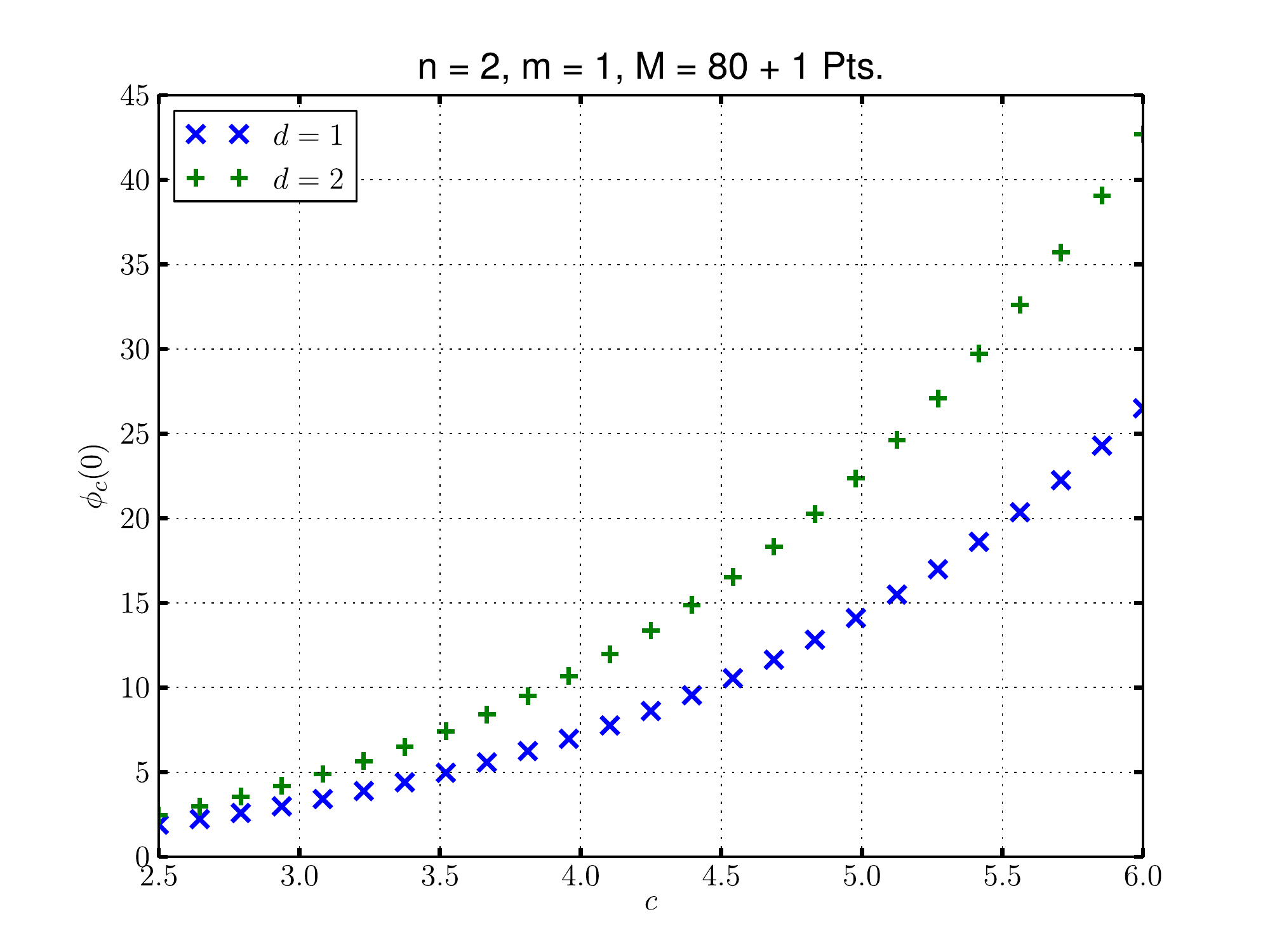}}
  \caption{Dispersion relations between $c$ and solitary wave
    amplitude.}
  \label{f:dispersion_relation}
\end{figure}

We can further consider this relationship between dimension and
amplitude by plotting ``dispersion relations'' between the parameter
$c$, and the amplitude of the associated solitary wave.  See Figure
\ref{f:dispersion_relation}.  The $n=2$, $m=1$ case has a much more
nonlinear dispersion relation than $n=3$, $m=0$ case.

In all cases with $m=1$, we did not show any $d=3$ solitary waves.
While performing numerical continuation in $d$ for these cases, our
solver failed to converge at an intermediary value of $d$ between 2
and 3.

\subsection{Convergence}
Not only does the sinc-collocation approach work, but it also
converges quite rapidly to the desired solution.  We now present some
benchmarks on the convergence of the amplitudes of the solitary waves
in different dimensions, for different choices of $(c,n,m)$.  We focus
on comparing the amplitudes ($\phi_c(0)$). Other points are more
difficult to compare as the grids change with $M$. The results are
given in Tables \ref{t:converge_c4_n3_m0}, \ref{t:converge_c5_n2_m1},
\ref{t:converge_c6_n4_m05}.

\begin{table}[p]
  \centering
  \caption{Convergence of the solitary wave amplitude for $c=4$ for
    the $n=3$, $m=0$ problem. }
  \label{t:converge_c4_n3_m0}
  \begin{tabular}{r l l l}
    \hline
    M & $d =1$ & $d=2$ &$d= 3$\\
    \hline
    20 & \underline{1.500}21353765 & \underline{1.70}417661440 & \underline{1.9}6849289246 \\
    40 & \underline{1.500000}80060 & \underline{1.706}08902282 & \underline{1.974}66312561 \\
    60 & \underline{1.50000000}989 & \underline{1.70617}046612 & \underline{1.9748}6670209 \\
    80 & \underline{1.500000000}23 & \underline{1.70617}690685 & \underline{1.97488}105982 \\
    100 & \underline{1.5000000000}1 & \underline{1.706177}67812 & \underline{1.974882}64950 \\
    200 & \underline{1.50000000000} & \underline{1.706177828}34 & \underline{1.974882937}68 \\
    400 & \underline{1.50000000000} & \underline{1.70617782848} & \underline{1.97488293789} \\
    800 & \underline{1.50000000000} & \underline{1.70617782848} & \underline{1.97488293789} \\
    \hline
  \end{tabular}
\end{table}

\begin{table}[p]
  \centering
  \caption{Convergence of the solitary wave amplitude for $c=5$ for
    the $n=2$, $m=1$ problem. }
  \label{t:converge_c5_n2_m1}
  \begin{tabular}{r l l}
    \hline
    M & $d =1$ & $d=2$\\
    \hline
    20 & \underline{14}.3312283238 & \underline{22}.5954364016 \\
    40 & \underline{14.2972}695906 & \underline{22.66}43001828 \\
    60 & \underline{14.297236}8619 &\underline{22.666}7057152 \\
    80 & \underline{14.29723672}60 & \underline{22.6668}188493 \\
    100 & \underline{14.2972367248} & \underline{22.66682}7544 \\
    200 & \underline{14.2972367248} & \underline{22.666828609}5 \\
    400 & \underline{14.2972367248} & \underline{22.666828609}6 \\
    \hline
  \end{tabular}
\end{table}

\begin{table}[p]
  \centering
  \caption{Convergence of the solitary wave amplitude for $c=6$ for
    the $n=4$, $m=.5$ problem.}
  \label{t:converge_c6_n4_m05}
  \begin{tabular}{r l l l}
    \hline
    M & $d =1$ & $d=2$ &$d= 3$\\
    \hline
    20 & \underline{1.479}45862654 &   \underline{1.6}7961148369 & \underline{1.9}4941026371 \\
    40 & \underline{1.479382}32695 &   \underline{1.680}59282799 & \underline{1.952}17168937 \\
    60 & \underline{1.47938214}568 &   \underline{1.68062}352158 & \underline{1.9522}3978609 \\
    80 & \underline{1.479382144}11 &   \underline{1.680625}57559 & \underline{1.95224}385885 \\
    100 & \underline{1.47938214408} & \underline{1.680625}79033 & \underline{1.952244}25280 \\
    200 & \underline{1.47938214408} & \underline{1.6806258265}3 & \underline{1.9522443147}3 \\
    400 & \underline{1.47938214408} & \underline{1.6806258265}5 & \underline{1.9522443147}6 \\
    \hline
  \end{tabular}
\end{table}

\section{Time Dependent Simulations}
\label{s:time}

\begin{figure}[htbp!]
  \centering

  \subfigure[]{\includegraphics[width=.9\textwidth,keepaspectratio,clip]{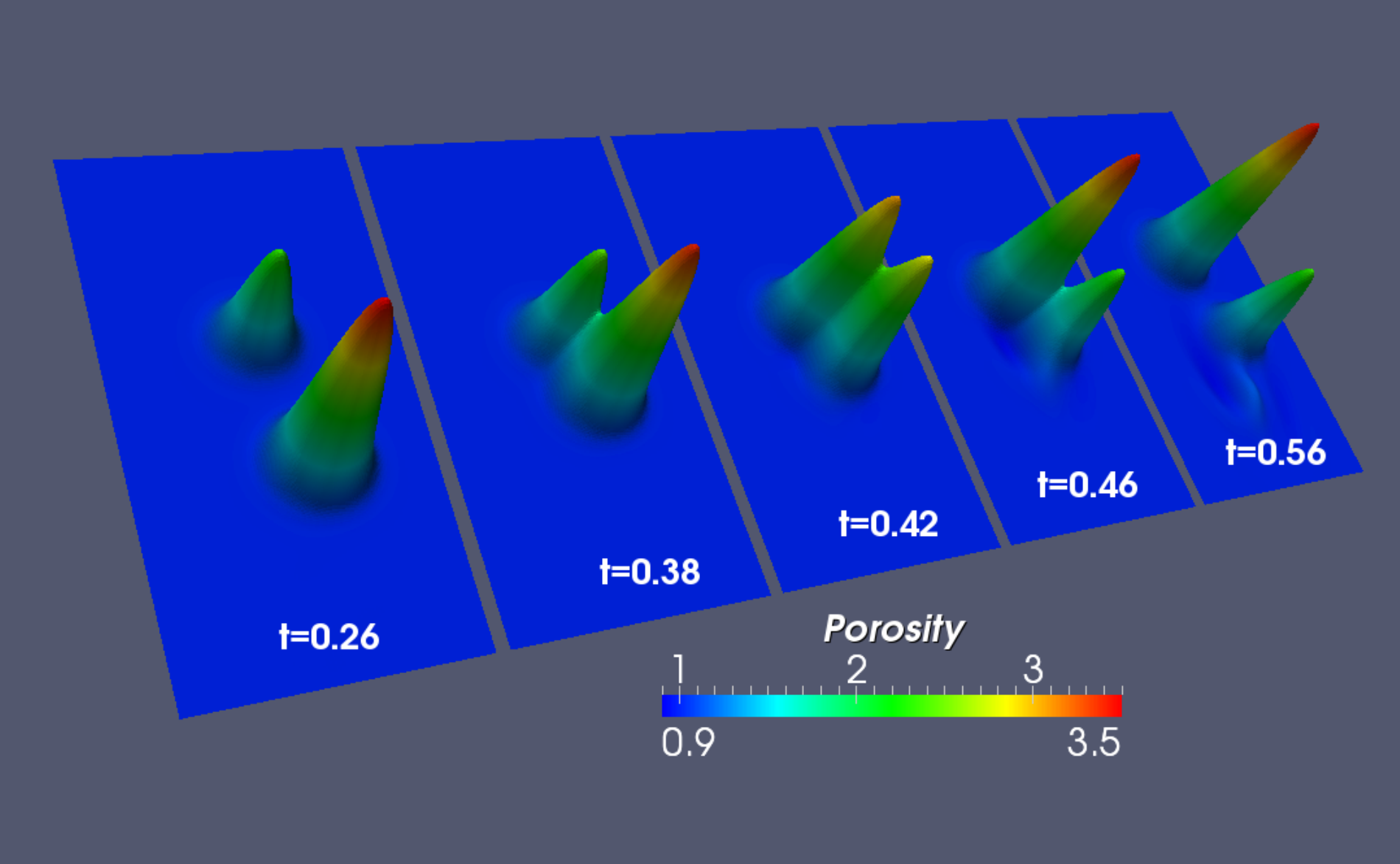}}
  \subfigure[]{\includegraphics[width=.9\textwidth,keepaspectratio,clip]{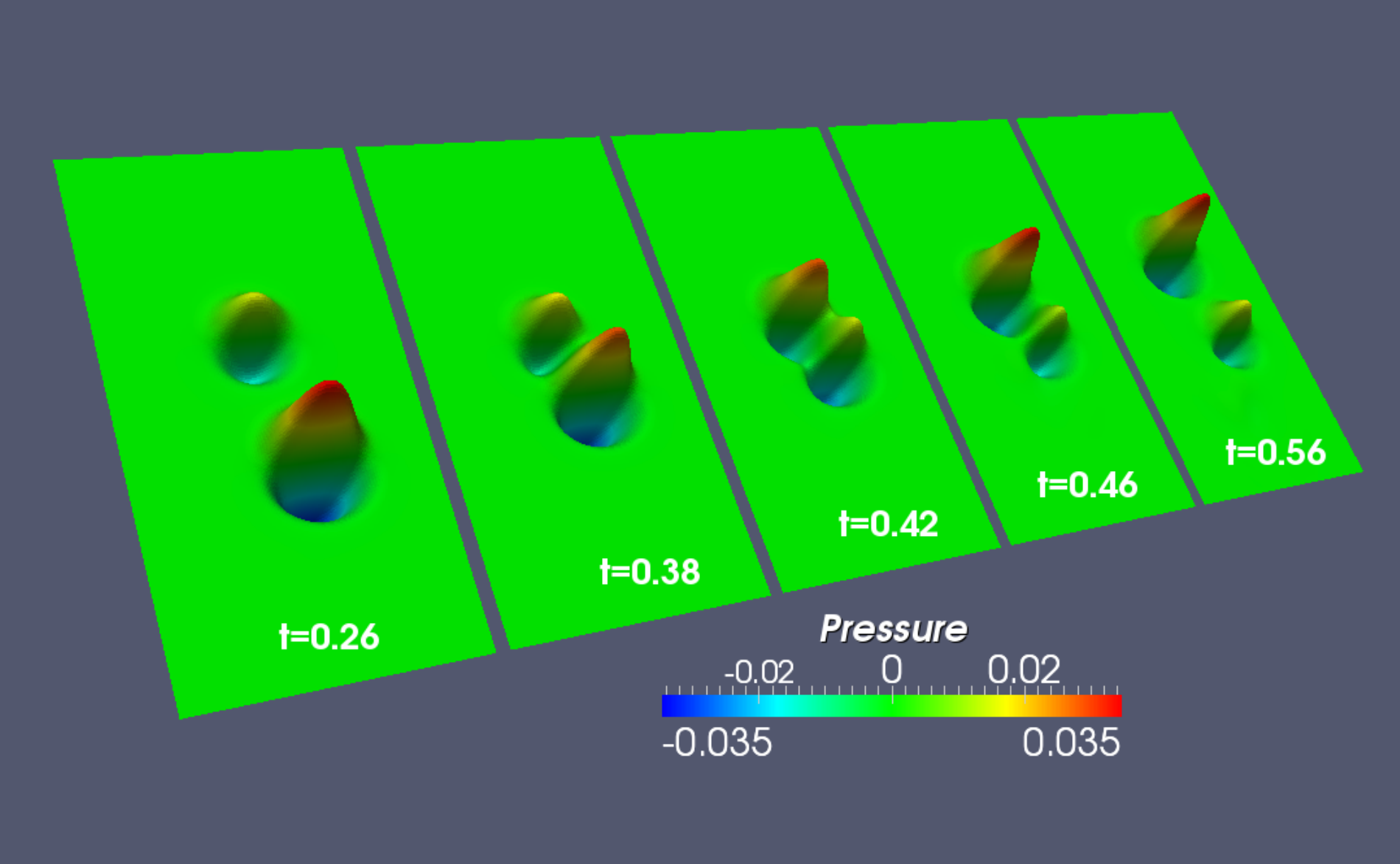}}

  \caption{Figure showing the off-center collision of two solitary
    waves with wave-speeds $c=7$ and $c=5$ and material exponents
    $n=3$, $m=0$. (\emph{a}) Porosity. (\emph{b}) Compaction
    Pressure. Model domain is $64\times128$ compaction lengths with
    $128\times256$ degrees of freedom (node spacing $h=0.5\delta$).
    Courant number is 1. This model is calculated in a frame moving at
    the mean speed of the two waves ($V=6$) and shows the typical
    non-linear phase shift interaction with some radiation loss on
    collision. }
  \label{fig:collision}
\end{figure}

Spectrally accurate solitary wave profiles are extremely useful as
initial conditions for benchmarking and exploring numerical solutions
of the full space time PDE's.  For example, Figure \ref{fig:collision}
shows a solution for an off-center collision of two 2-D solitary waves
with speeds $c=5$ and $c=7$ initialized with two sinc-collocation
solutions.  This calculation is done in a moving frame translating at
the mean velocity of the two waves which allows long runs in limited
numerical domains.  More specifically we solve a variation of the
coupled hyberbolic-elliptic problem, \eqref{e:magma_system} for
porosity $\phi$ and compaction pressure $\Pcmp$ \cite{Katz2007}:
\begin{subequations}\label{e:advective_magma_system}
  \begin{align}
    \DvD{}{\phi}{t} &=  \phi^m\Pcmp \\
    \left[-\Div\phi^n\Grad + \phi^m\right]\Pcmp &= -\nabla\cdot
    {\phi^n{\bf e}_d}
  \end{align}
  \label{eq:magma_fp}
\end{subequations}
where ${\bf e}_d$ is the unit vector in the direction of gravity and
\begin{displaymath}
  \DvD{}{\phi}{t} = \pdiff{\phi}{t} + \bv\cdot\Grad\phi
\end{displaymath}
is the material derivative in a frame moving at speed $\bv$ (which we
assume is constant for these problems, but can vary in space and time
for more general magma dynamics problems).

Given any numerical method for solving
\eqref{e:advective_magma_system}, the sinc solitary wave solutions
provide a straightforward benchmark problem: use a high resolution
($M=150$) solitary wave as an initial condition and solve. A perfect
scheme would have the wave propagate at speed $c$ with no change in
shape; anything else is numerical error.

\subsection{Numerical Methods} 
System \eqref{eq:magma_fp} has previously been solved by finite
difference and finite volume methods with explicit time stepping and
operator splitting \cite[e.g.]{wiggins1995mma}.  Here we describe and
benchmark more recent implicit finite element codes with
semi-Lagrangian/Crank-Nicolson time stepping for the advection terms.
Specifically, we solve the non-linear variational problem
\begin{equation}
  \label{eq:residual} 
  \begin{split}
    F(\bold{u}) & = \int_\Omega\left[ f^n\Grad v\cdot(\Grad p - {\bf
        e}_d) + v f^m p\right] dV
    +\int_{\partial\Omega} f^n{\bf e}_d\cdot d\bold{S}\\
    &  + \int_\Omega q (f -  \frac{\Delta t}{2} f^m p -g(\bold{x}^*)) dV = 0 \\
  \end{split}
\end{equation}
for consistent porosity, $f$, and pressure, $p$, at time $t+\Delta t$.
Here, $\bold{u}=(p,f)$ is a solution for pressure and porosity in a
mixed finite element space $\cal V$ with test functions $\bold{v} =
(v,q)$.  For the problems shown here we use second order elements on
triangular (2D) and tetrahedral meshes (3D) (i.e. ${\cal V} =
[P2\times P2]$).  The semi-Lagrangian source function
\begin{equation}
  \label{eq:1}
  g(\bold{x}^*) = f(\bold{x}^*,t) + \frac{\Delta t}{2}
  f(\bold{x}^*,t)^m p(\bold{x}^*,t),
\end{equation}
depends on the porosity and pressure at the previous time step
evaluated at the takeoff point of the characteristics that intersect
the quadrature points at time $t+\Delta t$
\cite[e.g.]{Staniforth91,Spiegelman2006}. For constant background
advection $\bold{x}^*=\bold{x} - \bv\Delta t$.

Equation \eqref{eq:residual} is non-linear with $F(\bold{u})$ being
the residual for any function $\bold{u}\in{\cal V}$.  We solve
$F(\bold{u})=0$, using pre-conditioned Newton-Krylov methods
implemented in hybrid \href{http://www.fenics.org}{FEniCS}
(http://www.fenics.org) and
\href{http://www.mcs.anl.gov/petsc/}{PETSc}
\cite{petsc-web-page,petsc-user-ref,petsc-efficient} codes.  FEniCS is
a suite of advanced, open-source software libraries and applications
that allows for high-level description of weak forms using a ``unified
form language'' (ufl) that can be translated into efficient,
compilable C++ code using their form compiler FFC.  In addition ufl
has the capability of describing and calculating the weak form of the
exact Jacobian ($J(u) = \delta F/\delta u$) by automatic functional
differentiation.  Given weak forms for both the residual and
Jacobians, FEniCS provides routines for assembly into discrete vectors
and matrices, which are solved using PETSc's non-linear equation
solvers (SNES).  These codes are highly flexible and can be used to
rapidly compose a wide range of PDE based models and adjust solver
strategies at run time.  We have used these codes to explore a variety
of physics-based block preconditioning strategies for iterative
solutions of the magma equations. Appendix \ref{sec:numerical-methods}
provides details of the numerical method used here.
Benchmark codes are available through the Computational Infrastructure
for Geodynamics (\href{http://www.geodynamics.org}{CIG}).

\subsection{Benchmark problems and results}

\begin{table}[tbp!]
  \centering
  \caption{Parameters for solitary wave benchmarks}
  \begin{tabular}{rrrrc}\hline
    $n$ & $m$ & $c$ & Amplitude & $\Omega$ (compaction lengths)\\\hline
    \multicolumn{5}{l}{\textbf{2D runs:}}\\ \hline
    3 & 0 & 5 & 2.33407 & $64\times64$ \\
    3 & 0 & 10 & 5.18711 & $64\times64$ \\
    2 & 1 & 2.5 & 2.44620 & $64\times64$ \\
    2 & 1 & 4 & 11.03790 & $64\times64$ \\ \hline
    \multicolumn{5}{l}{\textbf{3D runs:}}\\ \hline
    3 & 0 & 5 & 2.72588 & $32\times32\times32$ \\ \hline
  \end{tabular}
  \label{tab:benchark_parameters}
\end{table}

Table \ref{tab:benchark_parameters} gives parameters for four 2-D
problems and one 3-D problem.  For each problem we set the background
advection velocity to $\bold{v}= -c$ so that the wave appears
stationary in the moving frame. Each model run is calculated on a
square or cubic domain large enough so that the tails of the solitary
wave are $\phi_c-1<10^{-7}$ (see Table
\ref{tab:benchark_parameters}),with boundary conditions $(f,p)=(1,0)$
on the top edge and ``free-flux'' ($\Grad\Pcmp\cdot\hat{\bold{n}}=0$)
on the other three sides. For each wave, we consider a series of
spatial and temporal discretizations by varying the inter-node spacing
$h=.25$, $0.5$ and $1.0\delta_0$ where $\delta_0$ is the compaction
length in the constant porosity background.
Time steps are chosen such that each wave moves a fixed multiple of a
compaction length in a time-step i.e.  $c\Delta t/\delta_0 = 0.125$,
$0.25$, $0.5$, $1.0$ and $2.0$.  The Courant number is $c\Delta t/ h$.

Semi-Lagrangian schemes are characteristic based and do not have a CFL
stability condition. This allows for large time steps.  Nevertheless,
as the results here show, accuracy is degraded at large time-steps.
For this problem, the most efficient and accurate runs occur at
$\text{CFL}=1$.

For these problem, two natural measures of error are distortion of the
wave shape and disturbance of the phase speed.  Given a computed
solution for porosity $f(\bold{x},t)$, we identify both types of
errors by first minimizing the functional
\begin{equation}
  \mathcal{E}(\boldsymbol{\delta}) \equiv \int \paren{f({\bf x},t) -
    \boldsymbol{\phi}_c({\bf x} + \boldsymbol{\delta},t)}^2d {\bf x}
\end{equation}
where $\boldsymbol{\phi}_c$ is a high quality 
approximation of the solitary wave at time $t$.  Though this can be
done by direct minimization of the functional, it is more accurate to
solve the nonlinear system
\begin{equation}
  \inner{{{\bf f}({\bf x}) -
      \boldsymbol{\phi}_c({\bf x} + \boldsymbol{\delta})}}{\nabla \boldsymbol{\phi}_c({\bf x} + \boldsymbol{\delta}) }=0
\end{equation}

Numerical experimentation shows that this optimization/root finding
problem reduces to a single parameter system for the vertical
component of the displacement $\boldsymbol{\delta}$ due to the
symmetries of the equation. $\boldsymbol{\delta}$ measures the phase
shift, from which we can find the error in the speed parameter,
\[
\boldsymbol{\delta} \approx (\tilde{c} - c) t
\]
where $\tilde{c}$ is the speed of the numerically perturbed solitary
wave.  Once we have found $\boldsymbol{\delta}$, we can directly
compute the $L^2$ error in the approximations.

\begin{figure}[tbp!]
  \centering
  \subfigure[]{\includegraphics[width=.47\textwidth,keepaspectratio,clip]{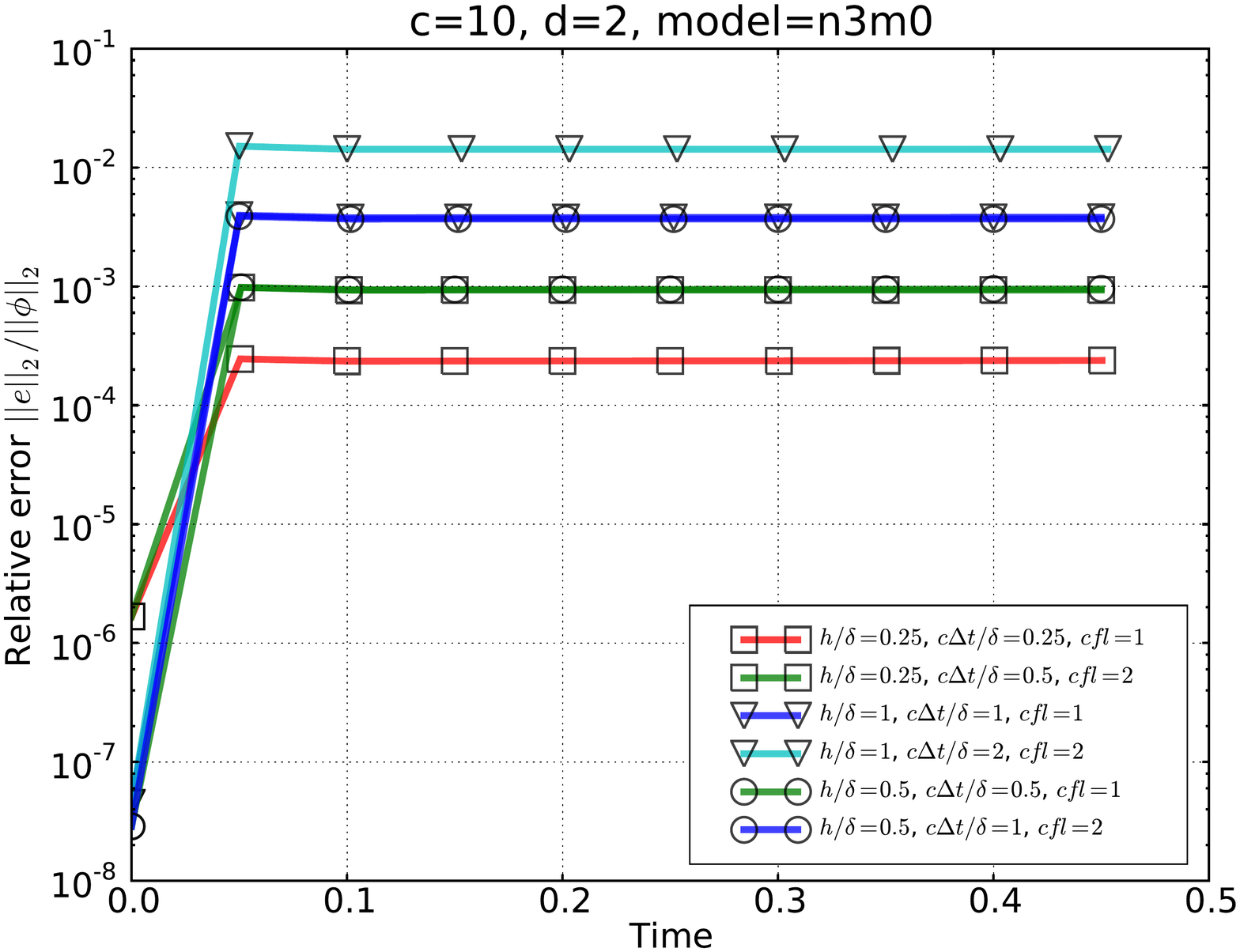}}
  \subfigure[]{\includegraphics[width=.47\textwidth,keepaspectratio,clip]{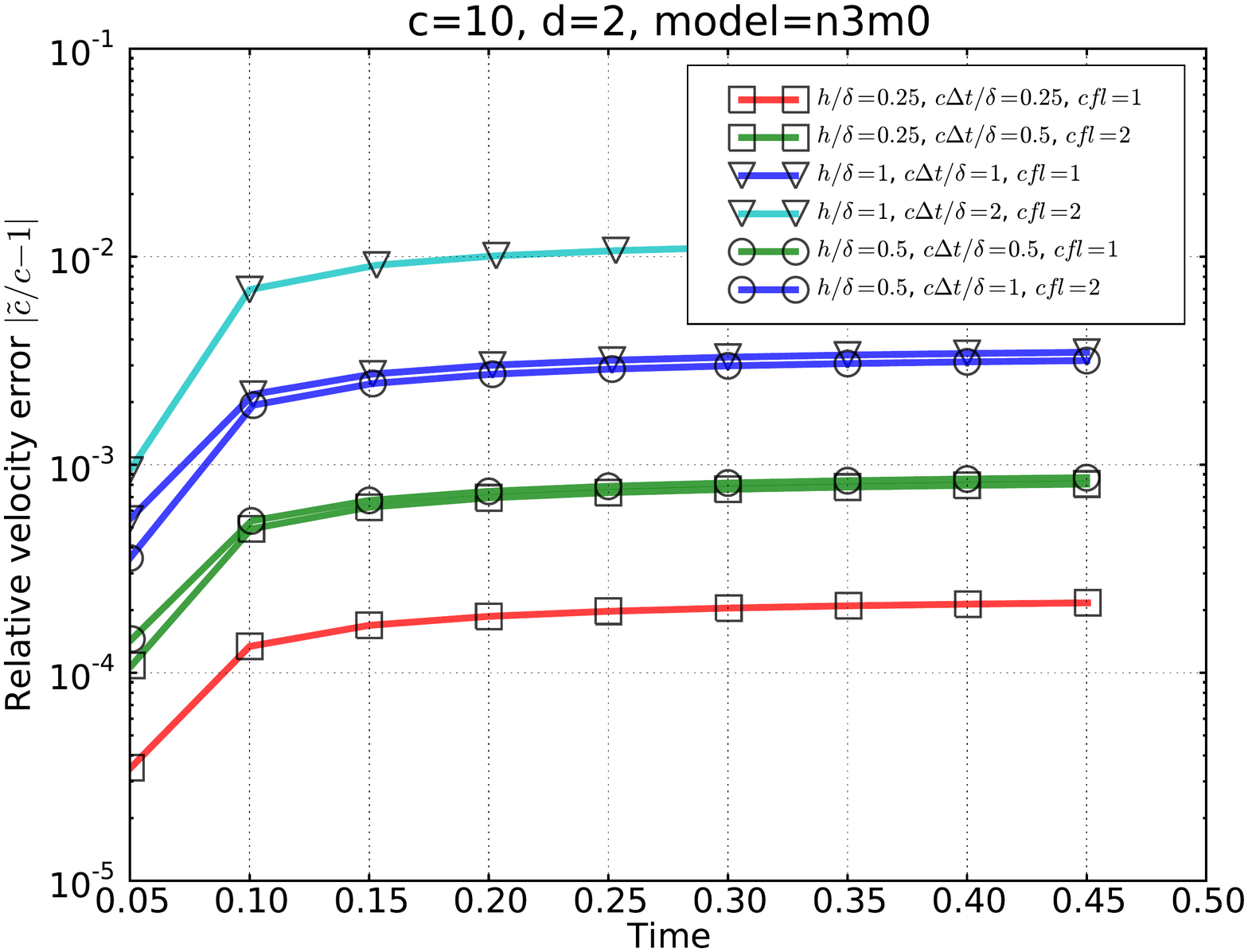}}
 \caption{Relative errors in (a) Shape and (b)
    Velocity for a 2-D, $n=3$, $m=0$, $c=10$ wave as a function of
    time, grid spacing $h/\delta_0$ and time-step $\Delta t$.  Grid spacing is
    the distance between nodes for quadratic elements and is relative
    to the compaction length in the constant porosity background.
    $c\Delta t/\delta_0$ gives the number of reference compaction lengths traveled in a
    time step.  The Courant number is $c\Delta t/h$.  For all runs,
    there is a rapid adjustment from the initial condition and then a
    steady evolution with little to no change in shape or
    velocity. Note, that the error depends almost entirely on the time
    step not courant number.  i.e. runs with the same time step and
    different courant numbers have very similar errors. 
  }
  \label{fig:errors_v_time}
\end{figure}

\begin{figure}[bpt!]
  \centering
  \subfigure[]{\includegraphics[width=.47\textwidth,keepaspectratio,clip]{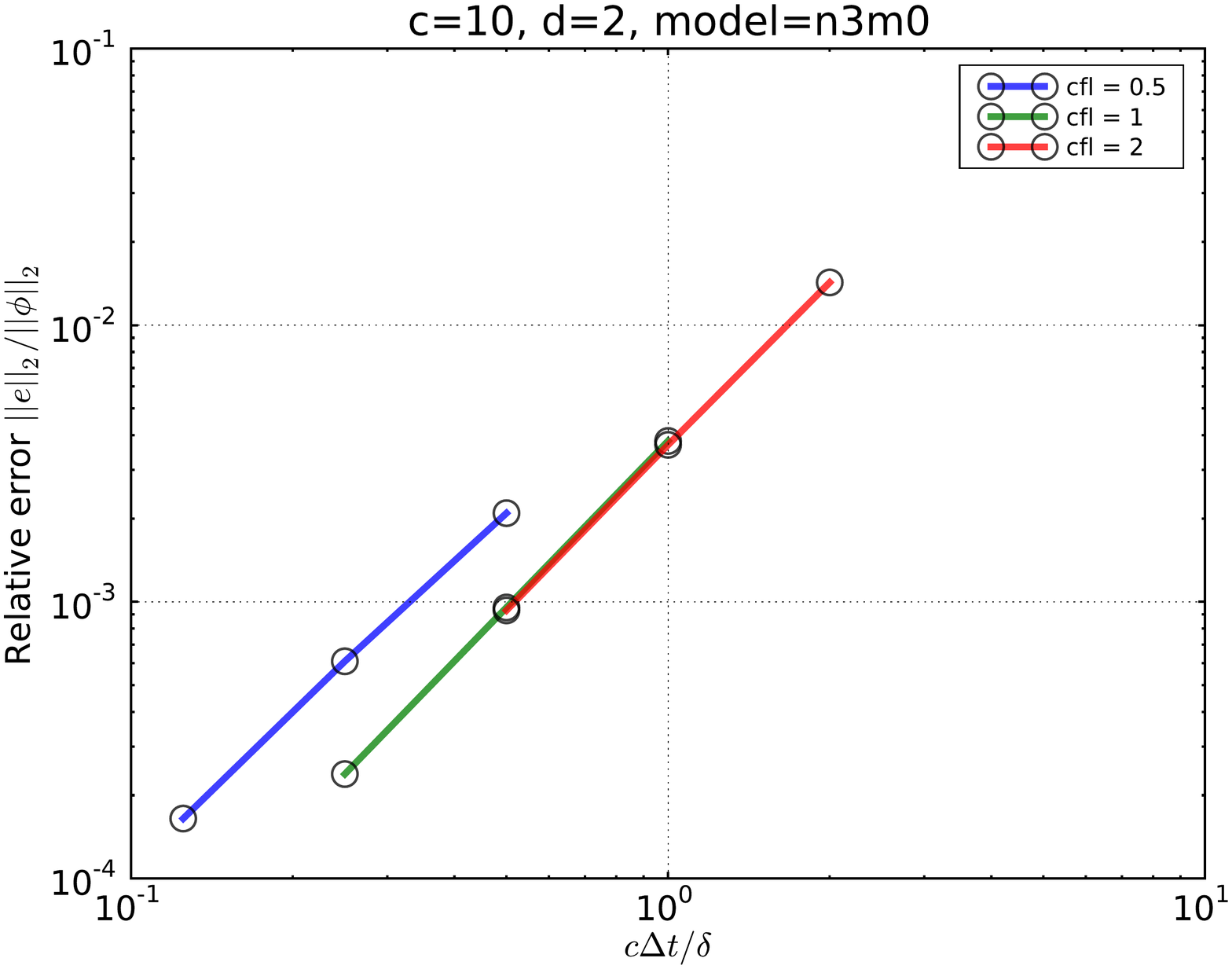}}
  \subfigure[]{\includegraphics[width=.47\textwidth,keepaspectratio,clip]{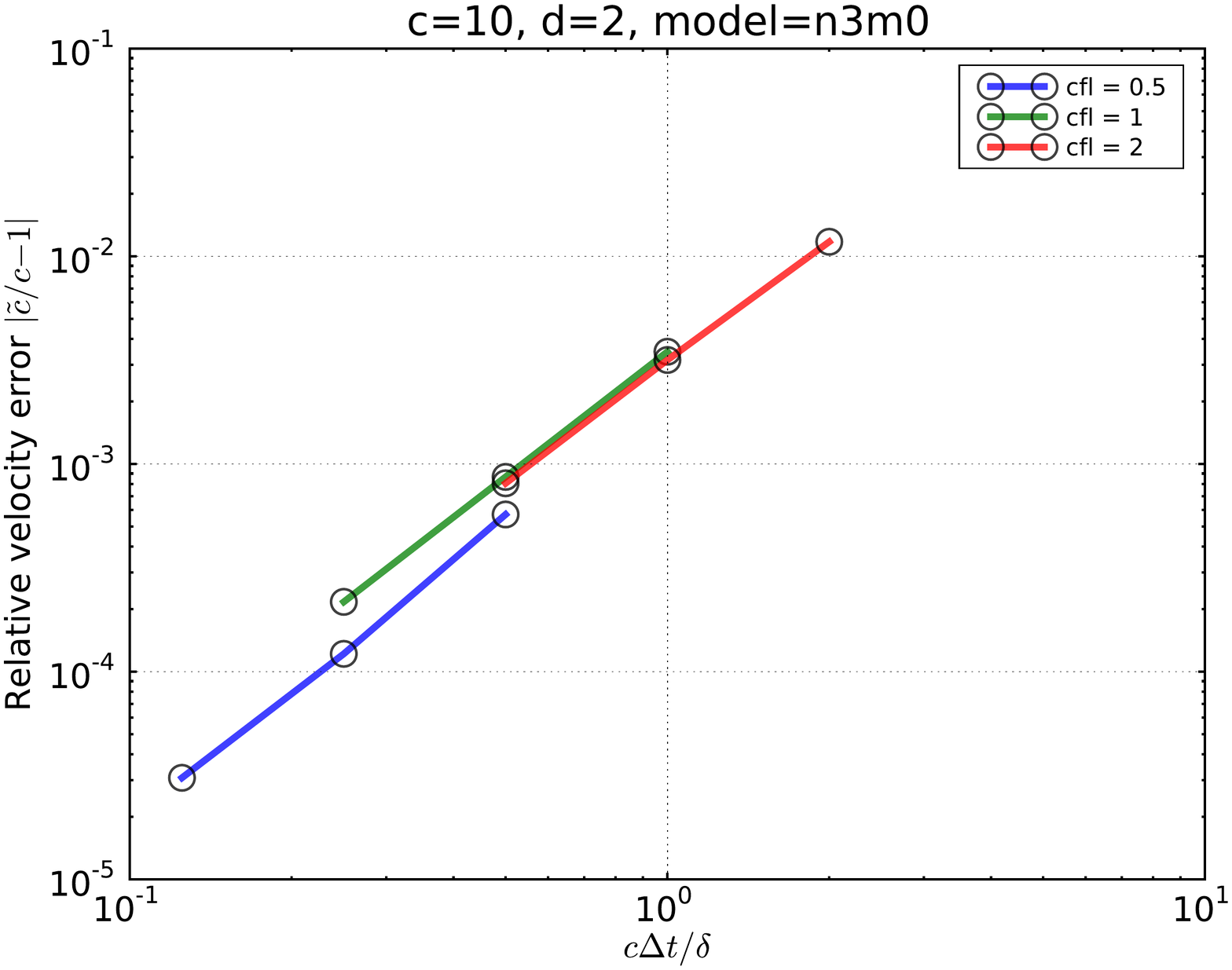}}

  \caption{Convergence behavior as a function of time step at $t\sim
    0.45$ for the same runs shown in Fig.\ \ref{fig:errors_v_time}.
    For any given courant number, the error shows second order
    convergence and runs with integer courant number show nearly
    identical errors for the same time-step $\Delta t$.  Runs with
    fractional courant number show similar convergence but larger
    shape errors (and in this case smaller velocity errors) at the
    same time-step.  For these runs highest accuracy per computational
    work occurs for courant number ${\rm cfl}=1$ runs.}
  \label{fig:cfl_figure}
\end{figure}

\begin{figure}[tbp!]
  \centering
  \includegraphics[width=.75\textwidth,keepaspectratio,clip]{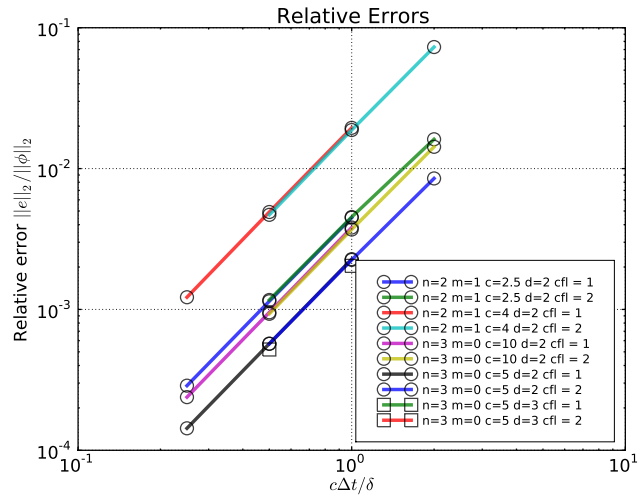}
  \caption{Convergence behavior for all runs as a function of
    time-step.  All problems show second-order convergence with
    $\Delta t$ and integer courant number.  For any given model,
    larger amplitude waves have larger errors and $n=2,m=1$ waves tend
    to have larger errors than $n=3,m=0$, likely due to the steeper
    porosity gradients in the $m=1$ waves. 3-D runs show similar
    errors to 2-D runs.  Relative error for all waves is $\lesssim
    10^{-3}$ for grid-spacing $h=0.25$, cfl=1.}
  \label{fig:errors_v_amplitude}
\end{figure}

Figure \ref{fig:errors_v_time} shows the relative shape error
$\sqrt{\mathcal{E}}/||\phi_c||_2$ and relative velocity error
$|\tilde{c}/c -1|$ as a function of time for a $n=3,m=0,c=10$ wave and
a range of grid spacing and time steps. Similar to previous solitary
wave studies \cite{Spiegelman1993b}, the initial conditions rapidly
adjust to an approximate solitary wave in the discrete function space
and then propagates with constant form and phase speed.  The numerical
methods shown here are second order as expected with errors that
depend primarily on the time step $O(\Delta t^2)$ (Figures
\ref{fig:cfl_figure}--\ref{fig:errors_v_amplitude}). Rough rules of
thumb for accuracy in these second order codes is that grid spacing
should be $h<0.25\delta_0$ with time step given by a courant number of
1.







\section{Discussion}
\label{s:discussion}

One of the challenges of studying Eq.\ \eqref{eq:magma} is that it is
fully nonlinear.  This carries over to the solitary wave equation
\eqref{eq:solwave_ode}.  As previously discussed, for $d>1$, this
cannot be reduced to a first integral.  For other multidimensional
equations, such as the Nonlinear Scrh\"odinger (NLS) equation and the
Zakharov-Kuznetsov (ZK) equation, one can apply spectral
renormalization/Petviashvilli's method
\cite{Ablowitz:2005p9,Lakoba:2007p151,Pelinovsky:2005p14}.  However,
these equations are semilinear, {\it i.e.}, they are already of the
form
\[
-\nabla^2 u + \lambda u - f(u) = 0.
\]
Thus, this popular approach is not immediately applicable to
\eqref{eq:solwave_ode}.

Another approach to solving for solitary wave profiles would be to use
a shooting algorithm, though it is highly unstable,
\cite{barcilon1989swm,wiggins1995mma}.  Other approaches are
comparable to our sinc collocation approach: use finite differences or
finite elements to construct a nonlinear algebraic system.  The
advantage of using sinc is that it naturally incorporates the boundary
conditions at infinity, whereas one would need to introduce an
artificial boundary condition for these problems.

It was necessary to extend $\f_c$ from $(0,\infty)$ to be a function
along the whole real axis to use sinc collocation.  We accomplished
this by an even extension; however, there are other possibilities.
The primary way of applying sinc methods to domains other than
$\mathbb{R}$ is to employ a function, $\psi(y)$, that maps
$\mathbb{R}$ into the domain of interest.  For the semi-axis,
suggested mappings are $\psi(y) = \log(y)$ and $\psi(y) = \log (
\mathrm{sinh}(y))$, \cite{lund1992smq}.  The main disadvantage to
mapping, as opposed to our even extension, is that we must modify our
scheme to accommodate the boundary condition $\f_c'|_{r=0} = 0$.


In general, the challenge of computing a solitary wave solution in
higher dimensions is not particular to \eqref{eq:magma}.  There is
similar difficulty with ZK and NLS and this method is likely to prove
very effective in these problems.

Beyond methods for accurate computation of non-linear waves, these
results provide the first critical tests for any code on magma
dynamics.  It should be stressed that Eqs.\
\eqref{e:advective_magma_system} represent the most simplified version
of magma dynamics that only include the contributions of non-linear
permeability and rheology to porosity evolution.  More general systems
are required to investigate the role of thermodynamics, chemistry, and
the interaction with the large scale mantle flow. However, at their
core, all of these problems need to reproduce the non-linear waves and
the benchmarks presented here are a necessary and reasonably
straightforward exercise in code verification. Moreover, these larger
systems, still have the general quasi-static structure of the basic
magma equations and require accurate and efficient multi-physics
solvers for coupled hyperbolic/parabolic/elliptic systems.  The
physics based, block-preconditioners demonstrated here for magma may
be a useful approach for more general problems.



\section*{Acknowledgements}
This work has been supported by NSF Grants OCE-08-41079 and
CMG-05-30853 and NSERC. The authors wish to thank M.
I. Weinstein for comments throughout the development of the sinc
approach.  G. Simpson wishes to also thank M. Pugh for suggestions and
C. Sulem for pointing out Sulem, Sulem \& Frisch, \cite{sulem1983tcs}, which was
instrumental in identifying the strip of analyticity.


\appendix
\section{Sinc Approximation}
\label{app:sinc_details}
Here we briefly review $\sinc$ and its properties.  The texts
\cite{lund1992smq, stenger1993nmb} and the articles
\cite{stenger1981nmb, bellomo1997nmp, stenger2000ssn} provide an
excellent overview.  As noted, sinc collocation and Galerkin schemes
have been used to solve a variety of partial differential equations.

\subsection{Overview}
Recall the definition of sinc,
\begin{equation}
  \label{eq:sinc_def}
  \sinc(z) \equiv \begin{cases}
    \frac{\sin(\pi z)}{\pi z}, &\text{if $z\neq 0$}\\
    1, &\text{if $z=0$}.
  \end{cases},
\end{equation}
and for any $k \in \mathbb{Z}$, $h>0$, let
\begin{equation}
  S(k,h)(x) = \sinc\paren{\frac{x-kh}{h}}.
\end{equation}

A sinc approximation is only appropriate for functions satisfying
certain criteria.  To define such functions we first define a
\emph{strip} about the real axis in the complex plane as
\begin{equation}
  \label{eq:stripdomain}
  D_\nu = \set{z\in \mathbb{C}\mid \abs{\Im z}<\nu} .
\end{equation}
Defining the function space:
\begin{defn}
  \label{def:bp}
  $B^p(D_\nu)$ is the set of analytic functions on $D_\nu$ satisfying:
  \begin{subequations}
    \begin{gather}
      \norm{f(t+ \mathrm{i} \cdot)}_{L^1(-\nu,\nu)} = O(\abs{t}^a),\quad \text{as $t\to \pm \infty$, with $a \in [0,1)$} ,\\
      \lim_{y\to \nu^-} \norm{f(\cdot+ \mathrm{i} y)}_{L^p}+
      \lim_{y\to \nu^-} \norm{f(\cdot- \mathrm{i} y)}_{L^p} <\infty .
    \end{gather}
  \end{subequations}
\end{defn}
we have the following
\begin{thm}[Theorem 2.16 of \cite{lund1992smq}]
  \label{thm:sinc_convergence}

  Assume $f \in B^p(D_\nu)$, $p=1,2$, and $f$ satisfies the decay
  estimate
  \begin{equation}
    \label{eq:decay}
    \abs{f(x)}\leq C \exp(-\alpha \abs{x}).
  \end{equation}
  If $h$ is selected such that
  \begin{equation}
    \label{eq:h_condition}
    h = \sqrt{\pi\nu/(\alpha M)}\leq \min\set{\pi \nu, \pi/\sqrt{2}} ,
  \end{equation}
  then
  \[
  \norm{\partial_x^{n} f -\partial_x^n C_M(f,h)}_{L^\infty}\leq C
  M^{(n+1)/2}\exp{\paren{(-\sqrt{\pi \nu\alpha M})}} .
  \]
\end{thm}
This theorem justifies the sinc method, and guarantees rapid
convergence for appropriate functions.  Checking that a function
satisfies all the hypotheses is non-trivial, and in practice it is
omitted.  However, it is essential to have a proper value of $h$ to
ensure convergence of algorithm; Theorem \ref{thm:sinc_convergence}
states the bound on $h$ is related to estimates of both the decay rate
and the domain of analyticity of the function.

\subsection{Decay Rate}
The decay rate is often easy to estimate.  For the solitary waves
solutions of \eqref{eq:solwave_ode}, the asymptotically linear
equation is
\[
-c \f_c ' + n \f_c' +c \paren{ \f_c''' + \frac{d-1}{r}\f_c'' -
  \frac{d-1}{r^2} \f_c'} = 0
\]
Integrating once,
\begin{equation}
  \label{eq:bessel}
  - \frac{c-n}{c} (\f_c -1) + \f_c'' + \frac{d-1}{r}\f_c'=0.
\end{equation}
The solutions in dimensions $n=1,2,3$ which decay as $r\to \infty$ are
\begin{equation*}
  \f_c-1 \propto \begin{cases} 
    e^{-\sqrt{1-n/c} r} & n=1,\\
    k_0(\sqrt{1-n/c} r) & n =2,\\
    \frac{1}{r} e^{-\sqrt{1-n/c} r} & n =3. 
  \end{cases}
\end{equation*}
$k_0$ is a Bessel function, and it decays exponentially.  Continuing
to assume that \eqref{eq:bessel} governs the large $r$ behavior, the
general decay relationship is
\begin{equation}
  \abs{\f_c(r)-1} \propto r^{-\frac{d-1}{2}}e^ {-\sqrt{1-n/c} r}
\end{equation}
The decay rate for Theorem \ref{thm:sinc_convergence} is thus $\alpha
= \sqrt{1-n/c}$.

\subsection{Analyticity}
\label{a:analyticity}
The second parameter, $\nu$, the distance into the complex plane which
the function can be analytically continued off the real axis, is not
as readily observed.  Others have found $\nu=\pi/2$ sufficient.  We
use $\nu = \pi/2$ and can numerically confirm \emph{a posteriori} that
this is a reasonable value.

Checking this condition is equivalent to identifying the decay rate in
Fourier space; $\nu$ satisfies
\begin{equation}
  \abs{\hat{u}(k)} \leq K e^{-\nu \abs{k}}
\end{equation}
To see this, write $u(x)$ in terms of its Fourier Transform,
\[
u(x) = \int_{-\infty}^\infty e^{i x k } \hat u (k) dk.
\]
Evaluating $u$ at $z = x + i y$,
\[
u(z) = \int_{-\infty}^\infty e^{i x k } e^{-yk} \hat u (k) dk.
\]
$u(z)$ is defined, provided
\[
\abs{u(z)} \leq \int_{-\infty}^\infty e^{-yk} \abs{u(k)}dk< \infty
\]
If $\hat u$ is bounded by $e^{-\nu \abs{k}}$, then this integral is
guaranteed to be finite for
\[
-\nu < y < \nu
\]
Thus, we can compute the Fourier transform and assess its decay rate
to identify $\nu$. This was used in \cite{sulem1983tcs} to study the
analyticity of the solutions to several partial differential
equations.

\begin{figure}
  \centering
  \subfigure{\includegraphics[width=2.2in]{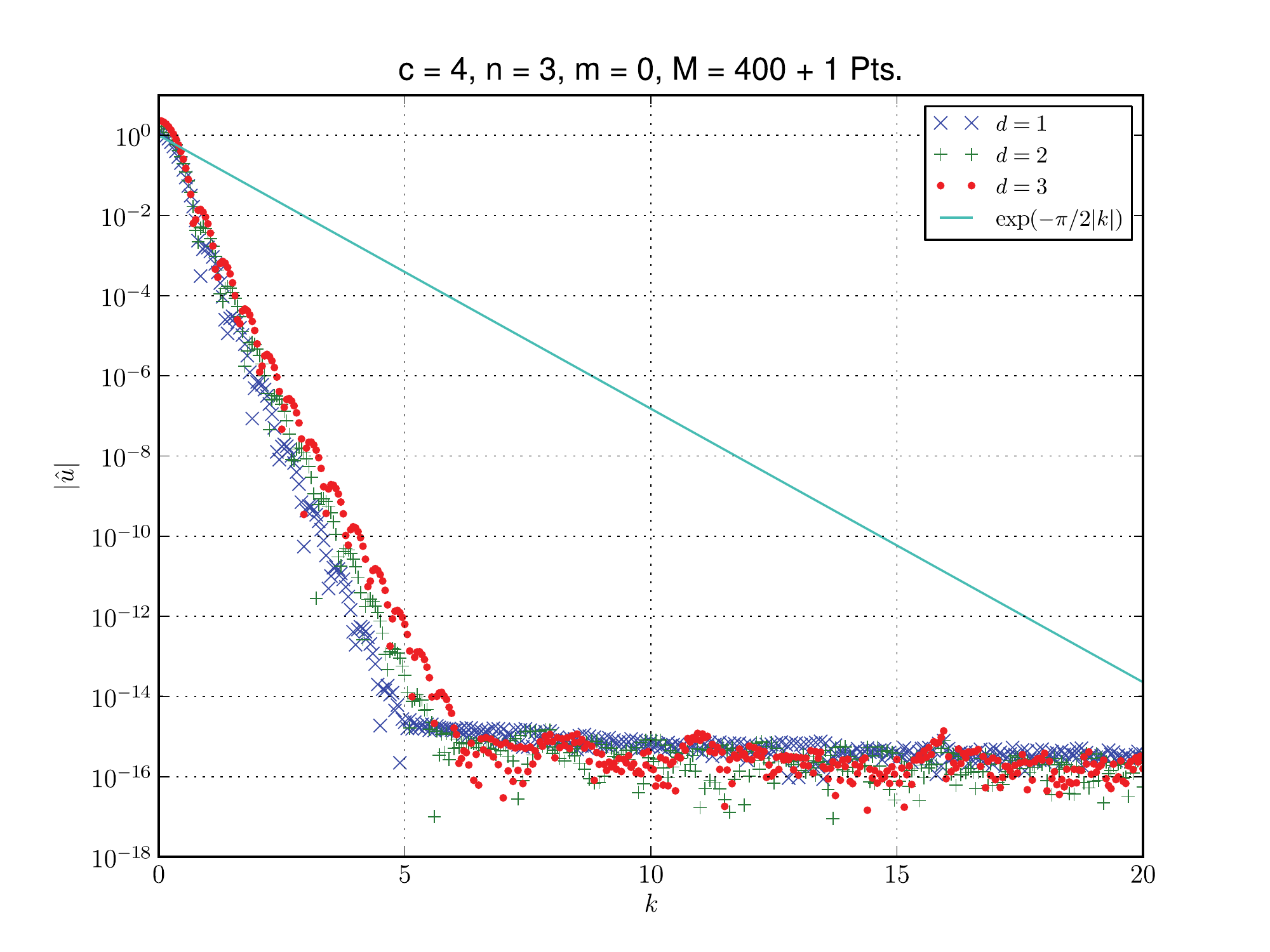}}
  \subfigure{\includegraphics[width=2.2in]{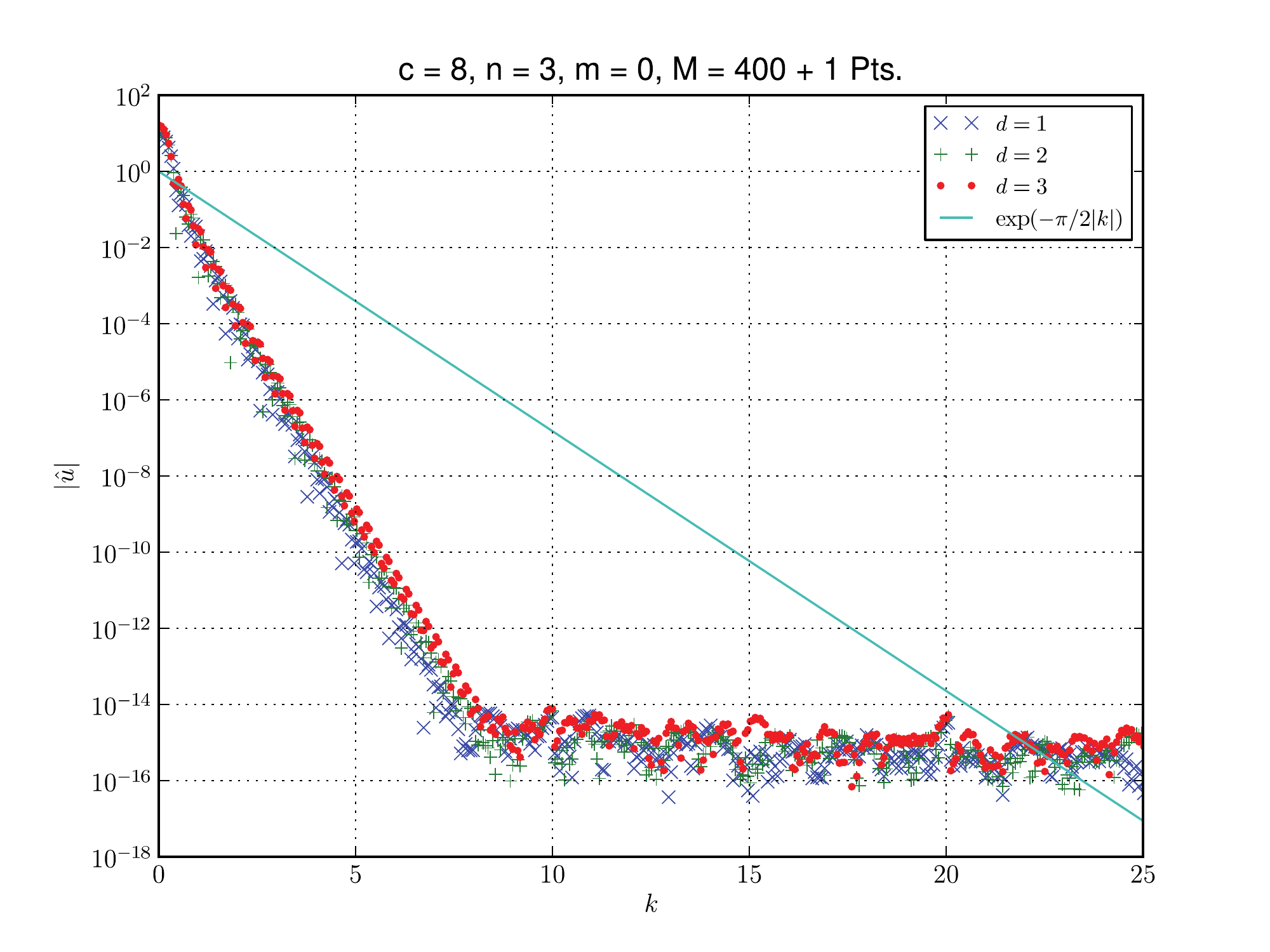}}

  \subfigure{\includegraphics[width=2.2in]{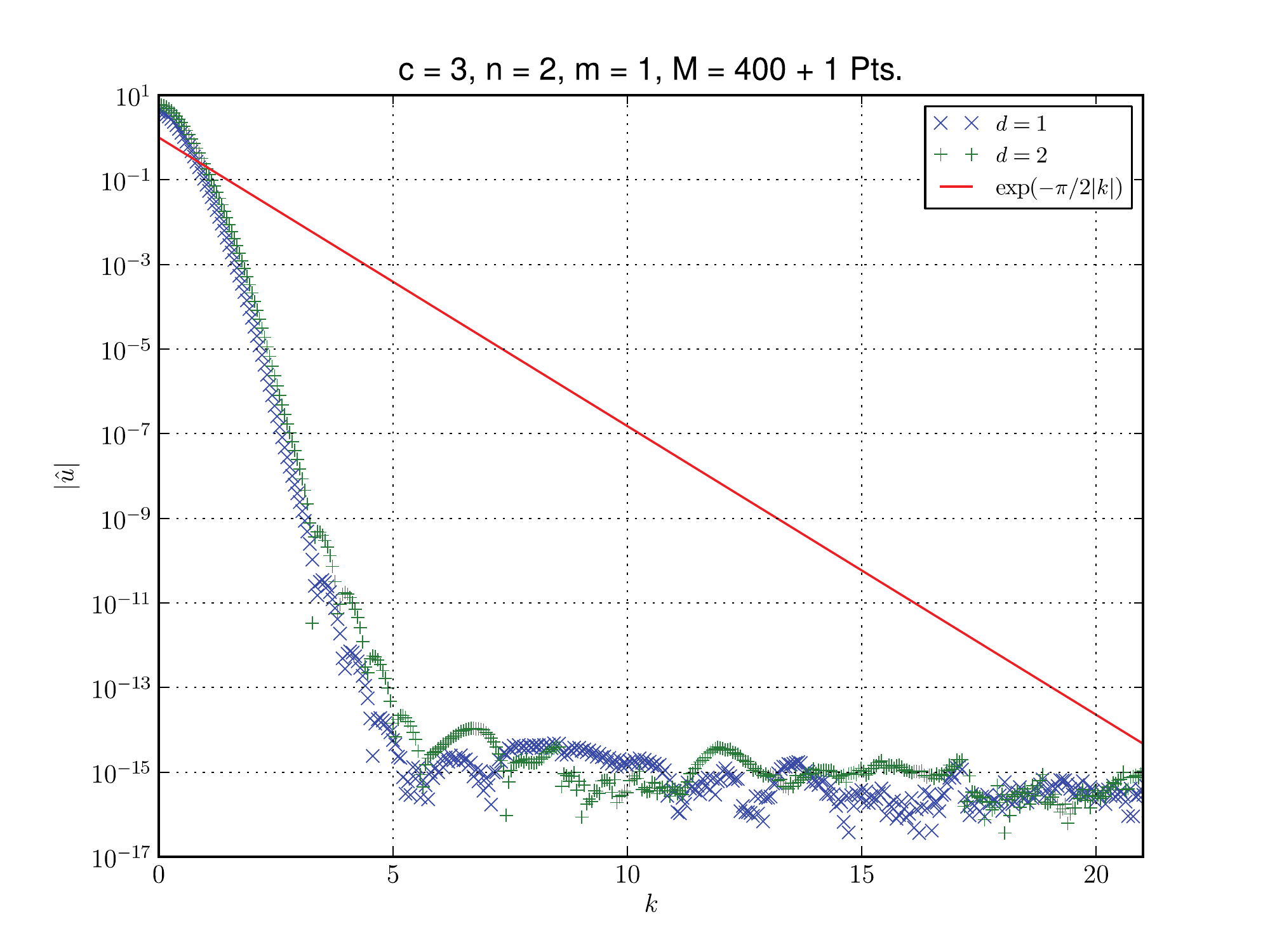}}

  \caption{Fourier transforms of the solitary waves computed by the
    sinc algorithm.}
  \label{f:ffts}
\end{figure}

Using our computed solitary waves, we approximate their Fourier
transforms as follows.  First, we form the even extension by
reflection.  Then we delete the node at $x_{-M}$; the solution now
sits on $2 M$ grid points.  This extended solution is treated as
periodic on $[x_{-M+1}, x_M]$ and its transform is computed.  Figure
\ref{f:ffts} shows the resulting transforms.  In these cases, we have
resolved them to machine precision.  More importantly, the solutions
decay more rapidly than than $e^{-\pi/2 \abs{k}}$.  This justifies
using $\nu = \pi /2$.  This value, together with $\alpha =
\sqrt{1-n/c}$, tells us that the largest value of $h$ for which we can
expect convergence is
\begin{equation*}
  h = \pi \sqrt{\frac{1}{2 \gamma M}}, \quad \gamma =  \sqrt{1-n/c}
\end{equation*}
which is \eqref{e:h_relation}.

\section{Notes on Numerical Methods for solution of space-time PDE's}
\label{sec:numerical-methods}

The algorithm demonstrated here uses second-order mixed Lagrange
finite elements (P2-P2) in space for porosity and pressure and a
second-order, semi-Lagrangian Crank-Nicolson scheme in time.  The
latter approach discretizes the generic hyperbolic advection-reaction
PDE
\begin{equation}
  \label{eq:2}
  \DvD{}{f}{t} = s(\bold{x},t)
\end{equation}
as a two level scheme via the trapezoidal rule
\begin{equation}
  \label{eq:3}
  f(\bold{x},t+\Delta t)=f(\bold{x}^*,t) + \frac{\Delta t}{2}
  \left[
    s(\bold{x},t+\Delta t) + s(\bold{x}^*,t)
  \right]
\end{equation}
where $\bold{x}^*$ is the take-off point of the characteristic that
intersects point $\bold{x}$ at time $t+\Delta t$
\cite{Staniforth91,Spiegelman2006}. We use this scheme to approximate
the strong form of Eq.\ (\ref{e:advective_magma_system}) and then
multiply by test functions and integrate to produce the non-linear
weak form of the residual at time $t+\Delta t$(Eq.\
\eqref{eq:residual}).

Semi-Lagrangian methods in finite elements can be considered a
distortion and reprojection problem.  For the pure advection
problem,$\DvDl{}{f}{t}=0$, the weak-form becomes
\begin{equation}
  \label{eq:semi-lag-pure-advection}
  \int_\Omega vf dx = \int_\Omega vf(\bold{x}^*) dx.
\end{equation}
This is just the projection of the advected continuous function
$f(\bold{x}^*,t)$ back onto the function space.  To evaluate the RHS
of Eq.\ \eqref{eq:semi-lag-pure-advection} by quadrature,
$f(\bold{x}^*)$ must be evaluated at the quadrature points. Therefore
$\bold{x}^*$ should be the coordinates of the take-off points of the
characteristics that intersect the quadrature point.  This is
different from finite difference problems where the characteristics
intersect the grid points.

Given the weak form of the residual, it is straightforward to
calculate the weak form of the Jacobian by functional differentiation.
This exact Jacobian is assembled by FEniCS into a $2\times2$
non-symmetric block linear problem
\begin{equation}
  \label{eq:Jacobian}
  \ttarray{A(f)}{B(f,p)}{\Delta t C(f)}{D(f,p)}\varray{\delta
    p}{\delta f}=-\varray{F_p}{F_f}
\end{equation}
where $\delta\bold{u}=[\delta p,\delta f]\trans$ is the correction to
the solution at each Newton step.  Note if $\Delta t=0$, the problem
becomes linear. If we begin with an initial guess $f=f_0, p=0$, then
the problem reduces to solving
\begin{equation}
  \label{eq:jacobian-dt0}
  \ttarray{A(f_0)}{B(f_0,p)}{0}{M}\varray{\delta
    p}{\delta f}=-\varray{F_p(f_0)}{0}
\end{equation}
where $M$ is the porosity mass matrix.  Thus for $\Delta t=0$, $\delta
f=0$ and $\delta p = -A(f_0)\inv F_p(f_0)$ which is just the discrete
solution to Eq.\ (\ref{e:advective_magma_system}b) for the pressure
given porosity $f_0$.

For a non-zero time-step, we solve the linear problem, Eq.\
(\ref{eq:Jacobian}) using a block-preconditioned Newton-Krylov scheme
implemented in PETSc, using their \texttt{FIELDSPLIT} block
preconditioners\footnote{PETSc gives considerable flexibility for
  experimenting with a wide range of solvers and pre-conditioners}.
Our preconditioner uses
\begin{equation}
  \label{eq:preconditioner}
  P=\ttarray{A}{0}{\Delta t C}{D}
\end{equation}
as a lower triangular block pre-conditioner with one V-cycle of
algebraic multi-grid on the $A$ block and 2 sweeps of SOR on the D
block, and then GMRES on the entire Jacobian.  Other choices of
preconditioners and solvers can be passed by command line
arguments. However, we find this particular recipe to be robust and
efficient in both 2 and 3-D, converging quadratically in two Newton
steps with total residuals $||F(\bold{u})||_2 < 10^{-14}$ independent
of dimension, grid-size, time-step and choice of model $(n,m)$ or
initial condition.  Using the pre-conditioner as a solver, turns the
algorithm into a Picard scheme with linear convergence (and a residual
reduction of about an order of magnitude per non-linear step).

Given a high quality initial condition for porosity generated by the
sinc-collocation scheme, we first project that solution onto our
initial porosity as $f_0$, the general time stepping algorithm is:
\begin{algorithm}
  \caption{SLCN-Newton-Krylov algorithm for
    Magma}\label{alg:full-newton-schemes-1}
  \begin{algorithmic}
    \REQUIRE at $t=0$, step $k=0$: Set initial condition $f_k=f_0$,
    $p_k=0$, \STATE set $\Delta t=0$, Solve for $p_k$ ($k=0$) by
    Newton \STATE set $\Delta t = dt$ \FOR[loop until $t \geq
    t_{max}$]{$k=1,2,\ldots$} \STATE Set initial Guess: \STATE
    \hspace*{2Em} set $p_k=p_{k-1}$ \STATE \hspace*{2Em} solve $Mf_k =
    \int q \left( g(\bold{x}^*) + \Delta t/2 f^m_{k-1}p_k \right) dx$
    \WHILE{$||F(\bold{u})||_2 >$ \texttt{tol} } \STATE Iterate
    preconditioned Newton-Krylov method for $f_k$,$p_k$
    \ENDWHILE
    \STATE $t \gets t+\Delta t$
    \ENDFOR
  \end{algorithmic}
\end{algorithm}

As with all non-linear solvers, having a good initial guess is
critical to robustness.  In the above algorithm we make a prediction
for the porosity at the future time by solving Eq.\
(\ref{e:advective_magma_system}a) as a projection problem with a
lagged pressure field.  $M$ is the symmetric porosity mass matrix
which can be solved with a few iterations of ICC preconditioned CG.

Though the above algorithm uses a constant time step, it is
straightforward to implement adaptive time stepping in these two-level
schemes for variable $\Delta t$.

\end{document}